\documentclass[%
 reprint, 
nofootinbib,
 amsmath,amssymb,
 aps, physrev,
]{revtex4-2}

\usepackage{graphicx,color}% Include figure files
\usepackage{dcolumn}% Align table columns on decimal point
\usepackage{bm}% bold math
\usepackage[final]{hyperref} % adds hyperlinks inside the generated pdf file
\usepackage{bbold}
\usepackage{graphicx}
\usepackage{caption}
\usepackage{subcaption}

\DeclareMathOperator{\Tr}{Tr}
\usepackage{bbold}

\begin{document}

\preprint{APS/123-QED}

\title{Fluctuation theorems for thermally isolated driven quantum systems: nonadiabaticity, excess work and strong inequalities}

\author{Jo\~ao V. M. Steimetz$^1$}
\email{Contact author: joao.steimetz@dac.unicamp.br}
\author{Michele Campisi$^2$}
\author{Marcus V. S. Bonança$^1$}%

\affiliation{%
$^1$Instituto de Física Gleb Wataghin, Universidade Estadual de Campinas, 13083-859 Campinas, São Paulo, Brazil}
\affiliation{$^2$NEST, Istituto Nanoscienze-CNR, Scuola Normale Superiore, I-56127 Pisa, Italy
}%

%\author{M. Campisi}
 %\homepage{http://www.Second.institution.edu/~Charlie.Author}
%\affiliation{second institution for this author}%

\date{\today}% It is always \today, today,
             %  but any date may be explicitly specified

\begin{abstract}
We expand on the ideas developed by C. Jarzynski in \textit{Physica A} \textbf{552}, 122077 (2020), where an integral fluctuation theorem was derived with the aim of obtaining thermodynamic inequalities stronger than those implied by the Jarzynski equality. Restricting ourselves to the quantum setting, we derive the corresponding detailed fluctuation theorem and additional detailed and integral fluctuation theorems; we also provide a clear physical interpretation of the stochastic quantities defined in the previous reference. Furthermore, we show that their averages are given by the nonadiabaticity parameter (i.e., the relative entropy between the final state after a finite-time driving protocol and the corresponding adiabatically evolved state) and the excess work (also known as inner friction). We elaborate on the inequalities derived from the fluctuation theorems and discuss their connection to irreversibility and formulations of the Second Law.
\end{abstract}

%\keywords{Suggested keywords}%Use showkeys class option if keyword
                              %display desired
\maketitle

%\tableofcontents

\section{\label{sec:intro}Introduction}

At the microscopic level, the quantities that describe a given system or process do not take on a sharp, unique value. Instead, they undergo fluctuations, and the value of a given quantity at a certain instant is characterized by a probability distribution.
The discovery of fluctuation relations \cite{Jarzynski1997_NW, Crooks1999_DetailedFT, CampisiHanggiTalkner2011_Review} is a major milestone in non-equilibrium statistical mechanics, as they clearly reveal a symmetry of the non-equilibrium probability distributions of the quantities that satisfy them. These relations follow directly from the initial distribution (usually taken as a Gibbs state) and the microscopic reversibility of the dynamics. 

Integral fluctuation theorems (IFT) usually lead to inequalities that represent statements of the Second Law. The most well-known example of this is the inequality, $W \geq \Delta F$, between the average work $W$ done on the system and the variation $\Delta F$ of the equilibrium free energy that follows from the Jarzynski equality \cite{Jarzynski1997_NW}
\begin{equation}
\label{eq:jarzynski_equality}
    \left\langle e^{-\beta(w-\Delta F)}\right\rangle = 1.
\end{equation}
Here, $w$ is the work done during a single realization of the process, so that $W = \langle w \rangle$. However, in the context of thermally isolated systems, it is known that this is not a particularly strong bound on the average work, in the sense that the bound cannot (in general) be saturated even for a quasistatic process \cite{Jarzynski2020, AllahverdyanNieuwenhuizen2005_MinimalWorkPrinciple}. As the Second Law of Thermodynamics implies the existence of a stronger bound on the work done on a thermally isolated system (see the Appendix of \cite{Jarzynski2020}), Jarzynski attempted in \cite{Jarzynski2020} to obtain it from fluctuation theorems. To this end, stochastic quantities were introduced and, although no clear physical interpretation could be given to them, an IFT was derived. However, the attempt to obtain the strong inequality from the IFT was only partially successful.

Proper thermodynamic inequalities that fully address irreversibility and the Second Law in the context of thermally isolated systems are desirable not only due to their fundamental character. They can also provide meaningful entropy production quantifiers when the system and its heat bath are strongly coupled \cite{Seifert2016,Jarzynski2017,Strasberg2017,Eisert2018,Talkner2020,Diba2024}. In this situation, usual notions such as the additivity of the thermodynamic entropy are lost \cite{Seifert2016,Jarzynski2017,Talkner2020,Strasberg2021}. However, the composite system may still be treated as a single thermally isolated one and the results corresponding to this scenario then apply.   

The outline of the article is as follows. In Section \ref{sec:theory}, we present a clear definition of the setting considered and the quantities involved in the subsequent results. In Section \ref{sec:X_DFT}, we show that, besides the IFT proved in \cite{Jarzynski2020}, there also exists the corresponding detailed fluctuation theorem (DFT). We prove it, interpret it, and connect it to previous results in the literature. In Section \ref{sec:Y_DFT}, we show that there are fluctuation theorems involving a second stochastic quantity. In Section \ref{sec:quantifiers}, we connect the two quantities featured in the fluctuation theorems to the nonadiabaticity parameter and the excess work, thus providing an interesting physical interpretation for both of them. In Section \ref{sec:irreversibility}, we connect their average values to irreversibility and the Second Law. As a consequence, one of the most interesting results of this paper is derived: a relation between the nonadiabaticity parameter and the work done on the system during what we call ``the cyclic counterpart'' of a given process. We also obtain directly from one of the IFT the inequality $W_\text{ex}^\text{th}\ge0$, where $W_\text{ex}^\text{th}$ is a quantity we term ``the thermodynamic excess work'', defined in Section \ref{sec:quantifiers}. This inequality is precisely the strong bound implied by the Second Law and mentioned before. We also discuss the difference between the excess work $W_\text{ex}$ and the thermodynamic excess work $W_\text{ex}^\text{th}$. Some calculations carried out in paradigmatic examples are presented in Section \ref{sec:examples} to illustrate our results. Conclusions and future avenues for research are discussed in Section \ref{sec:conclusions}.

\section{Theory and setting}
\label{sec:theory}

Next, we clearly define the situation and the quantities that lead to the desired fluctuation theorems. We stress once more that we restrict ourselves to thermally isolated quantum systems.

Let $\lambda_t$ be the value of the external parameter at time $t$. The set of values of $\lambda_t$ from the start of the process (at time $t=0$) to its end (at time $t=\tau$) specifies the driving protocol. Let $H(\lambda_t)$ be the Hamiltonian operator of the system at time $t$, with (possibly degenerate) eigenvalues $E_n(\lambda_t)$. The projector onto the eigenspace corresponding to $E_n(\lambda_t)$ is written as $P_n(\lambda_t)$, and the projectors onto different eigenspaces satisfy $P_n(\lambda_t)P_k(\lambda_t)=\delta_{nk}P_n(\lambda_t)$, as well as $\sum_nP_n(\lambda) = \mathbb{1}$. We assume that the degree of degeneracy of each eigenspace remains constant throughout the process. This excludes the possibility of degeneracy lifting and energy level crossings.

The Gibbs state (i.e., the thermal equilibrium state) for a certain inverse temperature $\beta=1/k_BT$ and a given value $\lambda$ for the external parameter is defined as 
\begin{equation}
    \label{eq:gibbs_def}
    \Pi_\beta(\lambda) = \frac{e^{-\beta H(\lambda)}}{Z_\beta(\lambda)} = \sum_n\mu_n(\lambda)P_n(\lambda),
\end{equation}
where $Z_\beta(\lambda) = \Tr\{ e^{-\beta H(\lambda)} \}$ is the partition function at inverse temperature $\beta$ and external parameter $\lambda$, while $\mu_n(\lambda) = e^{-\beta E_n(\lambda)}/Z_\beta(\lambda)$ are the weights corresponding to the canonical distribution.

The situation we consider from now on is analogous to that in which the Tasaki-Crooks fluctuation theorem applies. However, the stochastic quantities we define will not be directly related to the work, although they will be constructed from two energy measurements. We start our system of interest in the following initial density operator
\begin{equation}
    \rho(0)=\Pi_\beta(\lambda_0)\,,
\end{equation}
(due to possible initial thermal contact with a standard heat bath), and just after placing the system in thermal insulation, the first energy measurement is performed before any forward process is carried out. This first measurement of the two-point measurement scheme corresponds to a projective energy measurement of $H(\lambda_0)$ on $\rho(0)$, yielding energy $E_{n_i}(\lambda_0)$. Then, as $\lambda$ is varied from $\lambda_0$ to $\lambda_\tau$, the system evolves according to Hamiltonian dynamics, i.e., the evolution is dictated by the time evolution operator $U(t,0)$ satisfying $i\hbar \frac{d}{dt}U(t,0)=H(\lambda_t)U(t,0)$ and $U(0, 0) = \mathbb{1}$. At time $t=\tau$, a projective measurement of $H(\lambda_\tau)$ is carried out, and the energy $E_{n_f}(\lambda_\tau)$ is obtained.

\begin{figure}
    \centering
    \includegraphics[width=\linewidth]{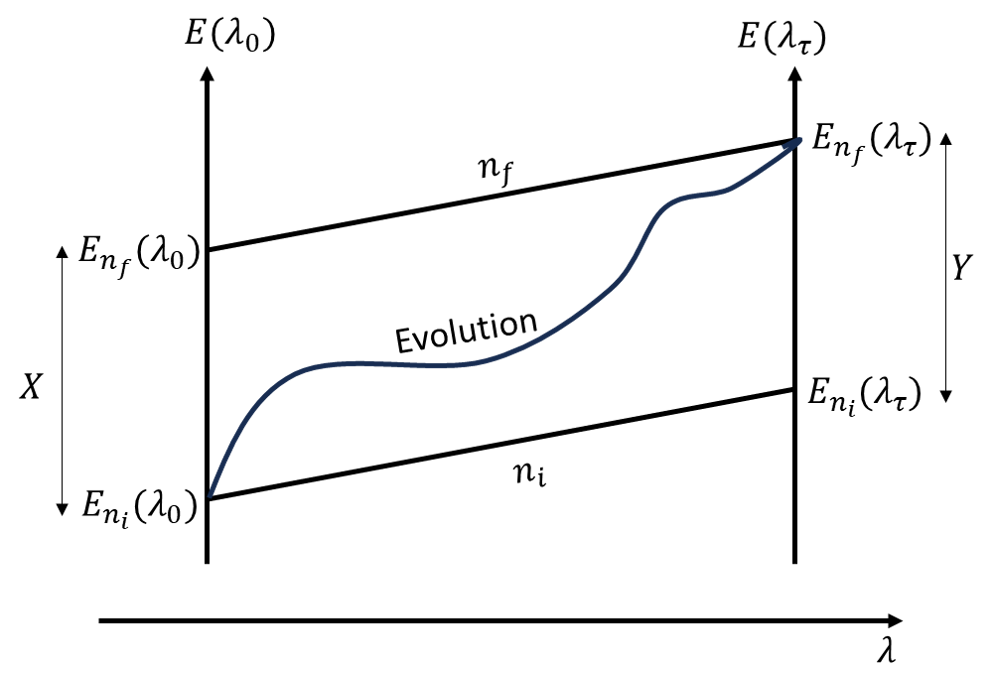}
    \caption{Illustration of the definition of $X$ and $Y$ in the quantum setting.}
    \label{fig:X_and_Y_def}
\end{figure}

Given this setup and the proposed measurements, in the context of thermally isolated driven systems, Jarzynski defined in \cite{Jarzynski2020} two stochastic quantities, named $X$ and $Y$ (due to the lack of a clear physical interpretation, we believe). They are given by
\begin{equation}
\label{eq:X_and_Y_def}
    \begin{split}
        X &\equiv E_{n_f}(\lambda_0) - E_{n_i}(\lambda_0) \equiv E_{n_fn_i}(\lambda_0)\\
        Y &\equiv E_{n_f}(\lambda_\tau) - E_{n_i}(\lambda_\tau) \equiv E_{n_fn_i}(\lambda_\tau).
    \end{split}
\end{equation}
These definitions are illustrated in Figure \ref{fig:X_and_Y_def}. Although there are four different energies involved, only two energy measurements are necessary (exactly those mentioned before) to determine all of them since $E_{n_f}(\lambda_0)$ is unambiguously related to $E_{n_f}(\lambda_\tau)$ once the energy spectrum is assumed to be known as a function of $\lambda$. The same applies to $E_{n_i}(\lambda_0)$ and $E_{n_i}(\lambda_\tau)$. 

We now introduce the quantities regarding the time-reversed process. They will be indicated by a tilde, and the argument or subscript accompanying them refers to the instant of time counted from the beginning of the time-reversed protocol. For instance, the initial density operator of the time-reversed process is denoted as $\tilde \rho(0)$, whose explicit form will be specified later. The values of the external parameter during the time-reversed protocol (denoted by $\tilde\lambda$) are given by $\tilde \lambda_{\tau-t} = \lambda_t$. That is, the external parameter assumes the same values as in the forward protocol, but in reverse order: the initial value of the external parameter in the time-reversed process corresponds to the final value in the forward process ($\tilde \lambda_0 = \lambda_\tau$) and vice versa ($\tilde \lambda_\tau = \lambda_0$). The Hamiltonian that governs the projective measurements and the evolution in the time-reversed process is $\Theta H(\tilde\lambda)\Theta^\dagger$, where $\Theta$ is the time reversal operator. Thus, the measurement of $E_n(\tilde \lambda_{\tau-t})$ in the time-reversed process is described by the projector
\begin{equation}
\tilde P_n(\tilde \lambda_{\tau-t}) = \Theta P_n(\lambda_t)\Theta^\dagger,
\end{equation}
while the evolution operator is given by 
\begin{equation}
\label{eq:micro_rev}
\tilde U(\tau-t,0) = \Theta U^\dagger(\tau, t)\Theta^\dagger;
\end{equation}
see App.~\ref{app:U_tilde} for the derivation.

For the remainder of this section, let us consider the case where the system's initial state $\rho(0)=\Pi_\beta(\lambda_0)$ evolves in the absence of projective energy measurements. This will be important from Sec.~\ref{sec:quantifiers} onward. In this setting, while the parameter varies from $\lambda_0$ to $\lambda_\tau$ according to the predefined protocol, the state of the system at time $t$ is
\begin{equation}
\label{eq:rho(t)_def}
\rho(t)\equiv U(t,0)\rho(0)U^\dagger(t,0).
\end{equation}
For a process of duration $\tau$, the state of the system at the end of the process is then $\rho(\tau) = U(\tau,0)\rho(0)U^\dagger(\tau,0)$.

One may then ask what the state of the system would be at the end of a very slow (in comparison to a typical time scale of the system) process that takes the parameter from $\lambda_0$ to $\lambda_\tau$.\footnote{In this work, $\lambda_0$ and $\lambda_\tau$ are fixed, chosen values, independent of the process duration. Thus, by increasing the process duration $\tau$, the variation in $\lambda$ becomes slower. The subindex $\tau$ in $\lambda_\tau$ only indicates that this is the value of $\lambda$ at the end of the protocol, not that $\lambda_\tau$ depends on $\tau$.} This is the subject of the Quantum Adiabatic Theorem, which we discuss in this paragraph, as the concept of adiabatic\footnote{In this text, the term ``adiabatic'' is used in its mechanical sense, i.e., ``adiabatic evolution'' means that the system evolves under infinitely slow variation of the control parameter. Here, it should not be interpreted in its thermodynamic sense, which refers to processes in which there is no heat exchange.} evolution will be crucial throughout this text. Although the Adiabatic Theorem is conventionally thought to apply only when the spectrum is nondegenerate at all times during a given process, there have been some works that lift this very restrictive condition, allowing degeneracies \cite{RigolinOrtiz2012} and even level crossings \cite{WangPlenio2016AdiabaticCondition, Xu2021Breaking}. Here, we treat the same case considered in \cite{RigolinOrtiz2012}: a possibly degenerate spectrum with degrees of degeneracy that are constant in time. In this case, one has \cite{RigolinOrtiz2010, RigolinOrtiz2012, RigolinOrtiz2014} 
\begin{equation}
\label{eq:ad_theorem}
    U_\text{ad}(t,0)P_n(\lambda_0)U_\text{ad}^\dagger(t,0) = P_n(\lambda_t),
\end{equation}
where $U_\text{ad}(t,0)$ corresponds to the time evolution operator that gives the adiabatic, transitionless evolution between the start of the process and the instant $t$. Since we take the initial state as $\rho(0)=\Pi_\beta(\lambda_0)$, the state obtained under adiabatic evolution up until the instant when $\lambda = \lambda_t$ is given by
\begin{equation}
    \label{eq:rho_ad def}
    \begin{split}
    \rho_\text{ad}(\lambda_t) &\equiv U_\text{ad}(t, 0)\Pi_\beta(\lambda_0)U_\text{ad}^\dagger(t,0)\\
    &= \sum_n \mu_n(\lambda_0)P_n(\lambda_t),
    \end{split}
\end{equation}
where we used Eqs. (\ref{eq:gibbs_def}) and (\ref{eq:ad_theorem}). By comparing Eqs. (\ref{eq:rho_ad def}) and (\ref{eq:gibbs_def}), one should note that the adiabatic state is generally not a Gibbs state. Therefore, even if the system starts in a Gibbs state and is driven infinitely slowly, it does not, in general, remain in a Gibbs state. This fact will be crucial throughout this text.

We are now in a position to derive fluctuation theorems for $X$ and $Y$ under the assumptions made thus far and to interpret the results. 

%%%%%%%%%%%%%%%%%%%%%%%%%%
\section{ First Fluctuation Theorem} 
\label{sec:X_DFT}

For completeness, we begin by proving the integral fluctuation theorem previously derived in \cite{Jarzynski2020}. However, we allow the spectrum to be degenerate, provided the degree of degeneracy of each eigenspace remains constant. The proof relies on the two-point measurement scheme and is as follows.

Let $p(m,n)$ be the joint probability of starting the two-point measurement scheme for the forward process in the eigenspace of $E_n(\lambda_0)$ and ending it in the eigenspace of $E_m(\lambda_\tau)$. Then,
\begin{equation}
\label{eq:IFT_derivation}
    \langle e^{-\beta X}\rangle = \sum_{m,n}p(m,n)e^{-\beta(E_m(\lambda_0)-E_n(\lambda_0))}.
\end{equation}
From the definitions in Sec.~\ref{sec:intro}, we may write 
\begin{equation}
    \label{eq:p(m,n) def}
    p(m,n) = \Tr\{ P_m(\lambda_\tau) U[P_n(\lambda_0)\rho(0)P_n(\lambda_0)]U^\dagger \}.
\end{equation}
Here, we have written $U\equiv U(\tau, 0)$ to simplify the notation. This convention will be used going forward, unless stated otherwise.

Besides, since $\rho(0)=\sum_k\mu_k(\lambda_0)P_k(\lambda_0)$, we have
\begin{equation}
\label{eq:p(m,n)}
    p(m,n) = \frac{e^{-\beta E_n(\lambda_0)}}{Z_\beta(\lambda_0)}\Tr\{ P_m(\lambda_\tau) UP_n(\lambda_0)U^\dagger \}.
\end{equation}
Substituting in Eq. (\ref{eq:IFT_derivation}) yields
\begin{equation}
\begin{split}
    \langle e^{-\beta X}\rangle &= \Tr\left\{\sum_m \frac{e^{-\beta E_m(\lambda_0)}}{Z_\beta(\lambda_0)} P_m(\lambda_0) U\sum_nP_n(\lambda_0) U^\dagger \right\}\\
    &= \Tr\left\{  \Pi_\beta(\lambda_0) U\mathbb{1} U^\dagger  \right\} = \Tr\{\Pi_\beta(\lambda_0)\} = 1,
\end{split}
\end{equation}
which is the desired result.

Before deriving the DFT for $X$, we should introduce a few more definitions. First, we take 
\begin{equation}
\tilde\rho(0) = \Theta \rho_\text{ad}(\lambda_\tau) \Theta^\dagger
\end{equation}
(in words, the time reversal of the adiabatic state at the end of the forward protocol) as the initial state in the time-reversed process. This is a rather unconventional choice of $\tilde \rho(0)$, as it is normally taken as a Gibbs state evaluated at $\lambda_\tau$. The physical reason for this choice shall be made clear in Sec.~\ref{sec:interpretation}. 

Furthermore, we must also define the random variable $\tilde X$ in the time-reversed process. We take 
\begin{equation}
\label{eq:X_tilde definition}
\tilde X = E_{\tilde n_f}(\lambda_0) - E_{\tilde n_i}(\lambda_0)
\end{equation}
or, in words, $\tilde X$ is defined as the difference between the energies regarding the quantum numbers obtained at the end and at the beginning of the time-reversed process, both evaluated at $\lambda_0$. Note that the energies are evaluated at the value of the parameter at the \textit{end} of the time-reversed process, a choice whose meaning will also be explained in Sec.~\ref{sec:interpretation}.

We are now ready to prove the DFT. Let $\tilde p(n,m)$ be the joint probability of obtaining $E_m(\tilde\lambda_0=\lambda_\tau)$ as the result of the first projective measurement of the time-reversed process and $E_n(\tilde\lambda_\tau = \lambda_0)$ as the result of the second. Then, we have:
\begin{equation}
\label{eq:p_tilde(n,m) def}
    \tilde p(n,m) = \Tr\{ \tilde P_n(\tilde\lambda_\tau) \tilde U [\tilde P_m(\tilde\lambda_0)\tilde\rho(0)\tilde P_m(\tilde\lambda_0)]\tilde U^\dagger \}
\end{equation}
where we used the convention $\tilde U \equiv \tilde U(\tau,0)$ to simplify the notation. By using $\tilde P_n(\tilde \lambda_{\tau-t}) = \Theta P_n(\lambda_t)\Theta^\dagger$, $\tilde U=\Theta U^\dagger\Theta^\dagger$, and $\Theta^\dagger\Theta=\mathbb{1}$, we write
\begin{equation}
    \tilde p(n,m) = \Tr\{ \Theta P_n(\lambda_0) U^\dagger P_m(\lambda_\tau) \rho_\text{ad}(\lambda_\tau) P_m(\lambda_\tau) U \Theta^\dagger \}.
\end{equation}
Equation (\ref{eq:rho_ad def}) and $P_k(\lambda_\tau)P_m(\lambda_\tau) = \delta_{km}P_m(\lambda_\tau)$ yield
\begin{equation}
    \tilde p(n,m) = \mu_m(\lambda_0)\Tr\{ \Theta P_n(\lambda_0)U^\dagger P_m(\lambda_\tau) U \Theta^\dagger  \}.
\end{equation}
Since $\Tr\{ \Theta A \Theta^\dagger \} = \Tr\{ A^\dagger \}$ for any trace-class operator $A$, we have
\begin{equation}
    \tilde p(n,m) = \mu_m(\lambda_0)\Tr\{ U^\dagger P_m(\lambda_\tau)UP_n(\lambda_0) \}.
\end{equation}
Finally, by applying the cyclic property of the trace and comparing the result with Eq. (\ref{eq:p(m,n)}), we find 
\begin{equation}
\label{eq:p(m,n) and p_tilde(n,m)}
    \tilde p(n,m) = e^{-\beta(E_m(\lambda_0) - E_n(\lambda_0))} p(m,n).
\end{equation}

Now, we write the probability distributions for $X$ in the forward and time-reversed processes.
\begin{align}
    \label{eq:P}p(X=x) &= \sum_{mn} \delta[ x - (E_m(\lambda_0) - E_n(\lambda_0)) ]p(m, n),\\
    \label{eq:P_tilde}  \tilde p(\tilde X =x) &= \sum_{mn} \delta[ x - (E_n(\lambda_0) - E_m(\lambda_0)) ]\tilde p(n, m).
\end{align}
Since the Dirac delta is an even function, one may write
\begin{equation}
    \tilde p(\tilde X=-x) = \sum_{mn} \delta[ x - (E_m(\lambda_0) - E_n(\lambda_0)) ]\tilde p(n, m)
\end{equation}
and using Eq. (\ref{eq:p(m,n) and p_tilde(n,m)})
\begin{equation}
\begin{split}
    \tilde p(\tilde X = -x) = \sum_{mn} \delta[ x - (E_m(\lambda_0) - E_n(\lambda_0)) ]p(m, n)\\
    \times e^{-\beta(E_m(\lambda_0)-E_n(\lambda_0))}.
\end{split}
\end{equation}
Now, because of the $\delta$-function, the nonzero terms in the summation will always be such that the exponent in the exponential is equal to $-\beta x$. Thus,
\begin{equation}
\begin{split}
    \tilde p(\tilde X = -x) &= e^{-\beta x}\sum_{mn} \delta[ x - (E_m(\lambda_0) - E_n(\lambda_0)) ]p(m, n)\\
    &= e^{-\beta x} p(X=x)
\end{split}
\end{equation}
where Eq. (\ref{eq:P}) was used in the last equality. We therefore conclude that 
\begin{equation}
    \label{eq:DFT for X}
    \frac{p(X=x)}{\tilde p(\tilde X=-x)} = e^{\beta x}
\end{equation}
which is the DFT for $X$.

We should briefly note that during the derivation of the DFT, the only property of $\Theta$ we used was the fact that it is an anti-unitary operator. Thus, analogous fluctuation theorems can be derived by using an arbitrary anti-unitary operator $K$ instead of $\Theta$ in all the definitions: the initial state of the time-reversed process would then be $K\rho_\text{ad}(\lambda_\tau)K^\dagger$ and the system would evolve according to the Hamiltonian $KH(\tilde\lambda)K^\dagger$. See \cite{Campisi2023FalseOnsager} for a more detailed explanation.

\subsection{Interpretation of the DFT}
\label{sec:interpretation}

\begin{figure}
    \centering
    \includegraphics[width=\linewidth]{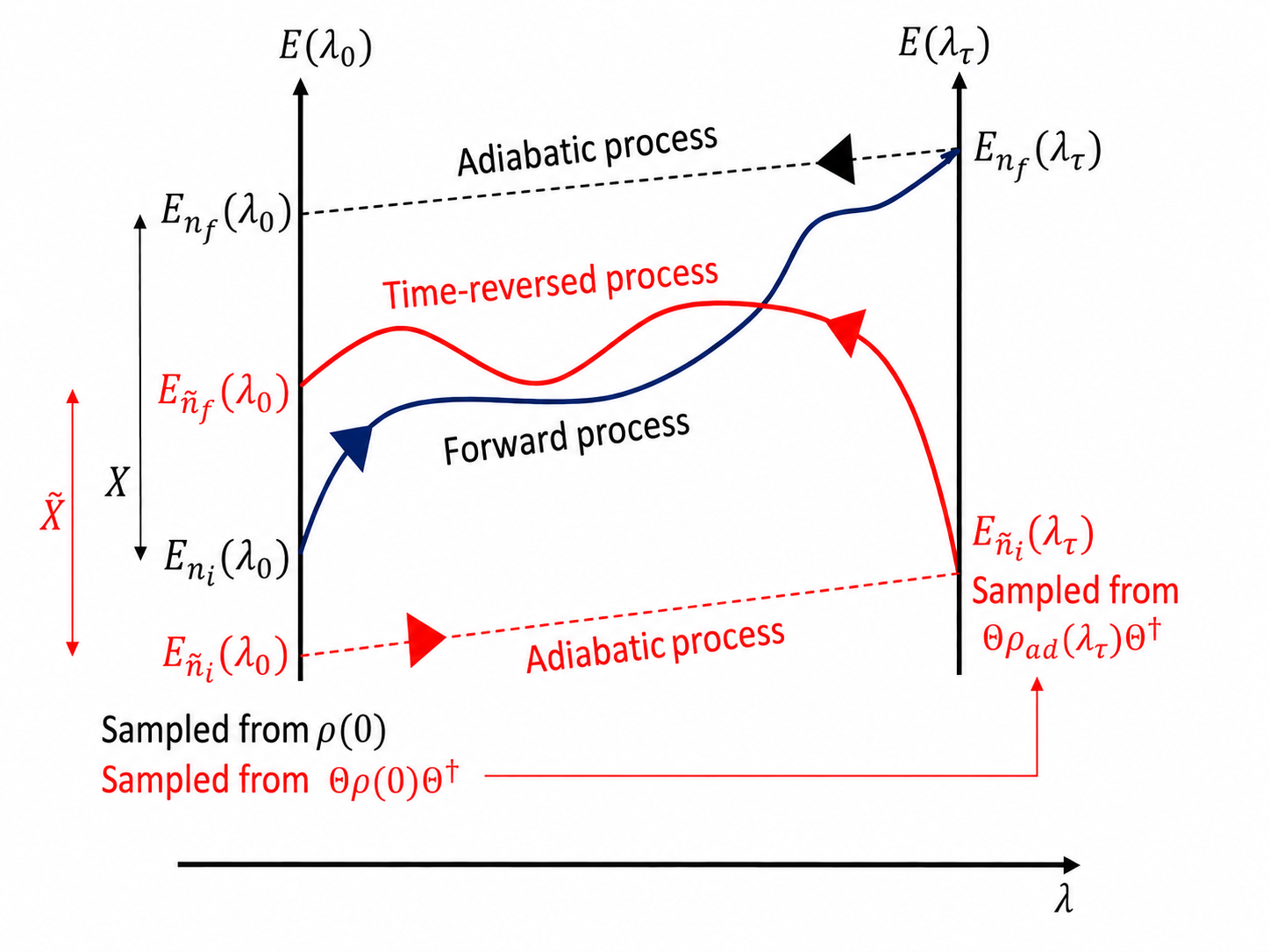}
    \caption{Illustration of the processes and quantities involved in the DFT.}
    \label{fig:X_DFT}
\end{figure}

We now provide an interpretation of the DFT that clarifies the physical meaning behind the choice of the initial state $\tilde \rho(0)$ and the definition of $\tilde X$ in the time-reversed process. An illustration of the explanation that follows is given in Figure \ref{fig:X_DFT}. 

First, we note that one may imagine that after the second projective measurement in the forward process, the driving parameter is brought back to its initial value $\lambda_0$ via an adiabatic process, thus creating an ``imaginary'' cyclic process\footnote{By cyclic, we mean that the final value of the external parameter is equal to its initial value. In this text, the word ``cyclic'' refers only to the value of the parameter and does not imply anything regarding the state of the system.}. Henceforth, we shall refer to this ``imaginary'' cycle as the ``cyclic counterpart'' of the ``original'' forward process or, alternatively, as the ``cyclic forward process''.

One then sees that $X$ for the forward process is just the work $w_c$ done during its cyclic counterpart for a single realization of the two-point measurement scheme. However, there is a subtlety that requires attention: it only makes sense to identify the energy variation $E_{n_f}(\lambda_0)-E_{n_i}(\lambda_0)$ with the mechanical work done during the imaginary cycle if there are no intermediate measurements in the course of the cycle. Yet, we have defined $X$ in terms of measurements taken at the beginning and the end of the \textit{original} forward process; i.e., there would be an intermediate measurement in the cyclic process. Regardless, one can easily prove that the random variables $X$ (obtained through the original definition of the two-point measurement scheme, that is, with measurements taken at the beginning and the end of the original process) and $w_c$ (the work done during the imaginary cycle if the measurements are taken at the beginning and the end \textit{of the cycle}, without intermediate measurements) are equal, $X=w_c$. This is due to the fact that the adiabatic portion of the imaginary cycle performs a one-to-one mapping between the eigenspace of quantum number $n$ at the beginning of the imaginary backward protocol and at its end. The proof of $X=w_c$ is straightforward and is provided in App.~\ref{app:X=w_c}.

Next, one takes the cyclic counterpart of the time-reversed process as the time reversal of the cyclic forward process. By the same reasoning developed in the previous paragraph, one concludes that $\tilde X = \tilde w_c$. We now understand the physical reason behind the definition in Eq. (\ref{eq:X_tilde definition}): by making the very reasonable choice to take the ``imaginary'' cycle of the time-reversed process as the time-reversal of the cyclic forward process, the definition in Eq. (\ref{eq:X_tilde definition}) is such that $\tilde X$ is identified with the work done during this time-reversed cycle. The choice of the cyclic counterpart of the time-reversed process also sheds light on the definition $\tilde \rho(0)=\Theta \rho_\text{ad}(\lambda_\tau)\Theta^\dagger$, since starting the original time-reversed process in the state $\Theta \rho_\text{ad}(\lambda_\tau)\Theta^\dagger$ is equivalent to starting its cyclic counterpart in the state $\Theta \rho(0)\Theta^\dagger$, the time-reversal of the initial state of the cyclic forward process.

We are now in a position to understand that the fluctuation theorems for $X$ are statements about the work done during the cyclic counterpart of a given process. These relations imply that arbitrary processes starting from thermal equilibrium are such that their cyclic counterparts tend to absorb energy instead of releasing it. More precisely, as one can easily show \cite{Jarzynski2020},
\begin{equation}
    p(w_c\le-x) = p(X\le -x) \le e^{-\beta x}.
\end{equation}
Thus, any process is such that its cyclic counterpart is exponentially unlikely to allow work extraction.

\subsection{Connection to previous results}
\label{sec:previous_results}

Once one sees that $X$ is equal to $w_c$, it becomes clear that the DFT for $X$ follows trivially from the Tasaki-Crooks fluctuation theorem \cite{Crooks1999_FT, Tasaki2000_QuantumJarzynski, TalknerHanggi2007_TasakiCrooks},
\begin{equation}
    \label{eq:tasaki-crooks}
    \frac{p(w=\omega)}{\tilde p(\tilde w = -\omega)} = e^{\beta(\omega - \Delta F)},
\end{equation}
where $w$ is the work done on the system during a single realization of the two-point measurement scheme in the forward process, $\tilde w$ is the analogous quantity for the time-reversed process, and 
\begin{equation}
\Delta F = -\beta^{-1}[\ln Z_\beta(\lambda_\tau)/Z_\beta(\lambda_0)]
\end{equation}
is the equilibrium free energy variation.
The initial distribution for the time-reversed process is, in this case, $\tilde \rho (0) = \Theta \Pi_\beta(\lambda_\tau)\Theta^\dagger$.

To see how Eq. (\ref{eq:DFT for X}) follows from Eq. (\ref{eq:tasaki-crooks}), let us apply the Tasaki-Crooks relation to a process consisting of the following steps:
\begin{enumerate}
    \item The system is initially in the state $\rho(0) = \Pi_\beta(\lambda_0)$, on which the first measurement of the scheme is performed. 
    \item A finite-time process is carried out with duration $\tau$, taking the external parameter from $\lambda_0$ to $\lambda_\tau$ according to the protocol $\lambda_t$. No measurement is performed between this step and the next one.
    \item The external parameter is taken back to $\lambda_0$ via adiabatic driving. At the end of this step, the second measurement of the scheme is performed.
\end{enumerate}
In order to apply the Tasaki-Crooks relation to this cyclic process, define the time-reversed process as follows:
\begin{enumerate}
    \item[$\tilde 1$.] The initial state is $\Theta \Pi_\beta(\lambda_0)\Theta^\dagger$. A projective measurement of $\Theta H(\lambda_0)\Theta^\dagger$ is performed.
    \item[$\tilde 2$.] An adiabatic process is carried out, taking the external parameter from $\lambda_0$ to $\lambda_\tau$ via quasistatic driving. No measurement is performed at the end of this step.
    \item[$\tilde 3$.] A finite-time process of duration $\tau$ is carried out, taking the external parameter from $\lambda_\tau$ to $\lambda_0$ according to the protocol $\tilde \lambda_{\tau - t} = \lambda_t$. The evolution is dictated by the Hamiltonian $\Theta H(\tilde\lambda)\Theta^\dagger$. At the end of this step, the second measurement is performed. 
\end{enumerate}
Of course, this is just the time reversal of the forward protocol described in steps 1-3. Since the process considered is cyclic, $\Delta F = 0$. It follows from the Tasaki-Crooks relation (\ref{eq:tasaki-crooks}) that 
\begin{equation}
    \frac{p(w_c=x)}{\tilde p(\tilde w_c = -x)} = e^{\beta x}.
\end{equation}
The equalities $X=w_c$ and $\tilde X = \tilde w_c$ introduced in Sec.~\ref{sec:interpretation} yield the DFT for $X$, Eq. (\ref{eq:DFT for X}).

Another way to derive the DFT for $X$ is to consider the forward and time-reversed processes just defined, but now performing energy measurements also at the end of steps 2 and $\tilde 2$. We then have a cyclic process with an intermediate measurement. This case can be treated according to the result of \cite{CampisiTalknerHaenggi2010_QuantumFluxes}, which states that the fluctuation theorem is not affected by intermediate projective measurements of any observable. Taking $w_c'$ (and $\tilde w_c'$ as the analogous quantity in the time-reversed process) as the energy difference between the measurements in steps 1 and 3, Eq. (\ref{eq:tasaki-crooks}) combined with the result of \cite{CampisiTalknerHaenggi2010_QuantumFluxes} yields
\begin{equation}
    \frac{p(w_c'=x)}{\tilde p(\tilde w_c' = -x)} = e^{\beta x}.
\end{equation}
One may easily see that $X=w_c'$ and $\tilde X = \tilde w_c'$, such that one recovers the DFT for $X$, Eq. (\ref{eq:DFT for X}).

%%%%%%%%%%%%%%%%%%%%%%%
\section{Second Fluctuation theorem}
\label{sec:Y_DFT}

In this section, we show an integral and a detailed fluctuation theorem involving $Y$, defined in Eq.~(\ref{eq:X_and_Y_def}). As we shall see, the IFT is just a different way of writing an existing IFT \cite{Tasaki2000_QuantumJarzynski}. However, the presentation developed here yields new insights and consequences, at least to the best of our knowledge. 

First, we must define the stochastic entropy at equilibrium. One usually defines the stochastic entropy at a given time as $-\ln p_r$, where $p_r$ is the probability that the system is in the microstate $r$ at that instant \cite{strasberg2022quantum}. Thus, we may define the stochastic entropy at equilibrium associated with the energy $E_n(\lambda)$ as $s_\beta(E_n;\lambda) = -\ln(\mu_n(\lambda)\Tr\{ P_n(\lambda) \}) = \beta E_n(\lambda) + \ln Z_\beta(\lambda) - \ln\Tr\{P_n(\lambda)\}$.

Now, let us consider an adiabatic process, where the quantum number remains constant. If one naively\footnote{Recall that the adiabatic state may not be a Gibbs state, as discussed in Sec.~\ref{sec:theory}.} takes the final state as a Gibbs state at temperature $1/k_B\beta'$, then one concludes that the stochastic entropy production associated with the initial quantum number $n_i$ is
\begin{equation}
\label{eq:sigma' def}
\begin{split}
    \sigma'&\equiv s_{\beta'}(E_{n_i};\lambda_\tau) - s_\beta(E_{n_i};\lambda_0)\\
    &= \beta'E_{n_i}(\lambda_\tau) - \beta E_{n_i}(\lambda_0) + \ln Z_{\beta'}(\lambda_\tau)/ Z_\beta(\lambda_0).
\end{split}
\end{equation}
As we have mentioned, the state obtained after adiabatic evolution may not be a Gibbs state. Indeed, we introduce the above quantity precisely as a measure of the violation of the expectation that the final state of an infinitely slow process is a Gibbs state. Let us justify this statement. 

First, one may easily see that calculating the average of $\sigma'$ with respect to the outcomes of the two-point measurement scheme yields\footnote{Note that, although the definition of $\sigma'$ was inspired by an adiabatic process, throughout the text we use the extension of this quantity to finite-time processes. This extension is formally exactly the same as Eq. (\ref{eq:sigma' def}); the only difference is that it can be evaluated for finite-time processes as well as adiabatic ones. It follows that, for a given initial quantum number, $\sigma'$ is the same (equal to the value calculated from Eq. (\ref{eq:sigma' def})) for all protocols from $\lambda_0$ to $\lambda_\tau$ and for all process durations, as it does not depend on the final quantum number.}
\begin{equation}
\label{eq:overline beta aux}
\begin{split}
    \langle \sigma'\rangle &= \sum_{m,n}p(m,n) [s_{\beta'}(E_{n};\lambda_\tau) - s_\beta(E_{n};\lambda_0)]\\ 
    &=\beta'\Tr\{ \rho_\text{ad}(\lambda_\tau)H(\lambda_\tau) \} + \ln Z_{\beta'}(\lambda_\tau)\\
    & \quad \  + \Tr\{ \Pi_\beta(\lambda_0)\ln\Pi_\beta(\lambda_0) \} \,.
\end{split}
\end{equation}
To obtain the first term of the second equality, we used
\begin{equation}
\begin{split}
  \sum_m \Tr\{ P_m(\lambda_\tau)& U P_n(\lambda_0) U^\dagger\} =
  \Tr\{\mathbb{1} U P_n(\lambda_0)U^\dagger\}\\
  &= \Tr\{ P_n(\lambda_0) \} = \Tr\{ P_n(\lambda_\tau) \}\,,
  \label{eq:degeneracy}
\end{split}
\end{equation}
since we have assumed a constant degree of degeneracy for each eigenenergy. Then, using $\beta' H(\lambda_\tau) = -\ln\Pi_{\beta'}(\lambda_\tau) - \ln Z_{\beta'}(\lambda_\tau)$ and  $\Tr\{ \Pi_\beta(\lambda_0)\ln\Pi_\beta(\lambda_0) \}=\Tr\{ \rho_\text{ad}(\lambda_\tau)\ln\rho_\text{ad}(\lambda_\tau) \}$, we find 
\begin{equation}
\label{eq:<sigma_bar> = D}
    \langle\sigma'\rangle = D[\rho_\text{ad}(\lambda_\tau)||\Pi_{\beta'}(\lambda_\tau)],
\end{equation}
where $D[\rho_1||\rho_2]\equiv \Tr\{ \rho_1\ln\rho_1\} - \Tr\{\rho_1\ln\rho_2 \}\ge0$ denotes the quantum relative entropy between states $\rho_1$ and $\rho_2$  \cite{nielsen2011quantum}. The quantum relative entropy $D[\rho_1||\rho_2]$ measures how difficult it is to distinguish the state $\rho_1$ from the state $\rho_2$ \cite{Vedral2002_RelativeEntropyReview}. Therefore, $\langle\sigma'\rangle = D[\rho_\text{ad}(\lambda_\tau)||\Pi_{\beta'}(\lambda_\tau)]$ measures the difference between the prediction from quantum mechanics ($\rho_\text{ad}$) and the expectation that the state obtained under an infinitely slow process is a Gibbs state. This justifies the statement from the previous paragraph. 

With the definition from Eq. (\ref{eq:sigma' def}), the following IFT is valid:
\begin{equation}
\label{eq:Y IFT}
    \langle e^{-\beta'Y - \sigma'}\rangle = 1, \forall \beta'
\end{equation}
Since $\beta'Y + \sigma' = \beta'E_{n_f}(\lambda_\tau) -\beta E_{n_i}(\lambda_0) + \ln Z_{\beta'}(\lambda_\tau) -\ln Z_\beta(\lambda_0)$, the IFT above is the same as the one derived in \cite{Tasaki2000_QuantumJarzynski}. For completeness, we provide the proof here as well:
\begin{equation}
\begin{split}
    \langle e^{-\beta'Y - \sigma'}\rangle &= \sum_{m,n}p(m,n)e^{-\beta'E_m(\lambda_\tau)+\beta E_n(\lambda_0)}\frac{Z_\beta(\lambda_0)}{Z_{\beta'}(\lambda_\tau)}\\
    &= \sum_{m,n}\Tr\{ P_m(\lambda_\tau)UP_n(\lambda_0)U^\dagger \} \frac{e^{-\beta'E_m(\lambda_\tau)}}{Z_{\beta'}(\lambda_\tau)}\\
    &= \sum_m\Tr\{ P_m(\lambda_\tau) \}\frac{e^{-\beta'E_m(\lambda_\tau)}}{Z_{\beta'}(\lambda_\tau)} = 1.
\end{split}
\end{equation}
As mentioned earlier, this is valid for any $\beta'$. By taking $\beta'=\beta$, we recover the Jarzynski equality, Eq. (\ref{eq:jarzynski_equality}).

By using appropriate definitions, we can also show a DFT involving $Y$. Let us first define the dimensionless random variable 
\begin{equation}
\label{eq:Y' def}
Y'\equiv \beta'Y+\sigma'.
\end{equation}
The definitions for the quantities in the time-reversed process are as follows:
\begin{align}
    &\tilde\rho(0) = \Theta \Pi_{\beta'}(\lambda_\tau)\Theta^\dagger\label{eq:rho_tilde DFT Y def}, \\
    &\tilde Y = E_{\tilde n_f}(\lambda_0) - E_{\tilde n_i}(\lambda_0),\\
    &\tilde\sigma' = \beta E_{\tilde n_i}(\lambda_0) - \beta'E_{\tilde n_i}(\lambda_\tau) + \ln Z_\beta(\lambda_0)/Z_{\beta'}(\lambda_\tau),\\
    &\tilde Y' = \beta \tilde Y + \tilde\sigma'.
\end{align}
The meaning behind these definitions can be understood by noting that, besides the standard time-reversal in Eq. (\ref{eq:rho_tilde DFT Y def}), the only changes in comparison to the definitions for the forward process were $\beta \leftrightarrow\beta'$ and $\lambda_\tau \leftrightarrow \lambda_0$. The latter is because the time-reversed process starts at the value $\lambda_\tau$ of the external parameter and ends at $\lambda_0$. The former is due to the fact that in Eqs. (\ref{eq:sigma' def}) and (\ref{eq:Y' def}), $\beta$ is the inverse temperature of the initial Gibbs state $\Pi_\beta(\lambda_0)$, while $\beta'$ refers to the temperature of a hypothetical Gibbs state $\Pi_{\beta'}(\lambda_\tau)$ at the end of the forward process; thus, when running the protocol in reverse, the change $\beta\leftrightarrow\beta'$ is justified.
In summary, we can say that the definitions above are completely analogous to the definitions for the forward process once we correctly identify the initial and final values of the temperature and the external parameter.

We are now in a position to prove the DFT. The proof is very similar to the one leading to the DFT for $X$, and so it will not be done in as much detail. To begin, one may easily prove that now the relation between $p(m,n)$ and $\tilde p(n,m)$ is\footnote{The relation is not the same as in Eq. (\ref{eq:p(m,n) and p_tilde(n,m)}) because the initial state of the time-reversed process is different.}
\begin{equation}
\label{eq:p_tilde DFT Y}
    \tilde p(n,m) = e^{\beta E_n(\lambda_0) - \beta'E_m(\lambda_\tau)}\frac{Z_\beta(\lambda_0)}{Z_{\beta'}(\lambda_\tau)}p(m,n).
\end{equation}
Then, we write the probability distributions
\begin{equation}
\begin{split}
    p(Y'=y') = \sum_{m,n}p(m,n)\delta[y'-&(  \beta'E_m(\lambda_\tau)-\beta E_n(\lambda_0)\\
    &+\ln Z_{\beta'}(\lambda_\tau)/ Z_{\beta}(\lambda_0))],
\end{split}
\end{equation}
\begin{equation}
\label{eq:p(tilde Y')}
\begin{split}
    \tilde p(\tilde Y'=y') = \sum_{m,n}\tilde p(n,m)\delta[y'-&(  \beta E_n(\lambda_0)-\beta' E_m(\lambda_\tau)\\
    &+\ln Z_{\beta}(\lambda_0)/ Z_{\beta'}(\lambda_\tau))].
\end{split}
\end{equation}
The substitution $y'\leftrightarrow-y'$ in Eq. (\ref{eq:p(tilde Y')}), followed by the use of the parity of the $\delta$-function and the substitution of Eq. (\ref{eq:p_tilde DFT Y}) into the resulting equation, yields the DFT for $Y'$:
\begin{equation}
\label{eq:Y_DFT}
    \frac{p(Y'=y')}{\tilde p(\tilde Y'=-y')} = e^{y'}.
\end{equation}

By taking $\beta'=\beta$, one recovers the Tasaki-Crooks DFT, Eq. (\ref{eq:tasaki-crooks}). The consequences of the fluctuation theorems derived in this section will be further explored in Sec.~\ref{sec:Y and minimal work}. 

%%%%%%%%%%%%%%%%%%%%%%%%%%%%%%%%
\section{Nonadiabaticity parameter and excess work}
\label{sec:quantifiers}

In this section, we show that the averages of  $X$ and $Y$ admit a clear physical interpretation by relating them to two well-known quantities: the excess work (or inner friction) and the nonadiabaticity parameter. Let us first present the definitions of these quantities. 

The nonadiabaticity parameter $\mathcal{A}$ is defined as 
\begin{equation}
    \mathcal{A}\equiv D[\rho(\tau)||\rho_\text{ad}(\lambda_\tau)].
\end{equation}
Here, $\rho(\tau)=U(\tau,0)\Pi_\beta(\lambda_0)U^\dagger(\tau,0)$ is the state of the system at the end of the finite-time process of duration $\tau$ in the absence of projective measurements, as defined in Sec.~\ref{sec:theory}. Besides, recall that the adiabatic state $\rho_\text{ad}(\lambda_\tau)$ is the limit of $\rho(\tau)$ as the protocol from $\lambda_0$ to $\lambda_\tau$ is carried out infinitely slowly.
Since the relative entropy measures the distinguishability between states \cite{Vedral2002_RelativeEntropyReview}, the nonadiabaticity parameter can be seen as a measure of adiabaticity breaking.

The excess work $W_\text{ex}$ is defined as 
\begin{equation}
\label{eq:W_ex def1}
    W_\text{ex} = W - W_\text{ad},
\end{equation}
where $W_\text{ad}$ is the average work done under adiabatic evolution and $W$ is the actual average work done during the finite-time process. Thus, the excess work quantifies the additional energy (in comparison to an adiabatic process) transferred to the thermally isolated system.

Under these definitions, we wish to show that 
\begin{equation}
\label{eq:X_and_Y_avgs}
    \beta\langle X\rangle = \mathcal{A} \quad \text{and} \quad \langle Y \rangle = W_\text{ex}.
\end{equation}

We begin by writing the averages of $X$ and $Y$ explicitly:
\begin{align}
        \label{eq:<X>}\langle X \rangle &= \sum_{m,n} p(m,n)E_{mn}(\lambda_0)\\
        \label{eq:<Y>}\langle Y \rangle &= \sum_{m,n} p(m,n)E_{mn}(\lambda_\tau).
\end{align}
Now, we manipulate the definitions of the nonadiabaticity parameter and the excess work to show that they are equal to $\beta\langle X\rangle$ and $\langle Y \rangle$, respectively. 

Let us start with the nonadiabaticity parameter. We write
\begin{equation}
\label{eq:nonadiabaticity_step1}
\begin{split}
\mathcal{A} &= \Tr\{ \rho(\tau)\ln\rho(\tau) - \rho(\tau)\ln\rho_\text{ad}(\lambda_\tau) \}\\
&= \Tr\{ \rho(0)\ln\rho(0) \} - \Tr\{ U\rho(0)U^\dagger U_\text{ad}\ln\rho(0)U_\text{ad}^\dagger \}   
\end{split}
\end{equation}
where we used the convention $U\equiv U(\tau,0)$ and $U_\text{ad}\equiv U_\text{ad}(\tau, 0)$ to simplify the notation, as well as the fact that $\ln(U'\rho'U'^\dagger)=U'\ln(\rho')U'^\dagger$, for all $\rho'$ and all unitary $U'$. Now, let us calculate each term in the last line of Eq. (\ref{eq:nonadiabaticity_step1}), starting with the first one.
From $\rho(0)=\Pi_\beta(\lambda_0)$, we get $\ln\rho(0) = -\beta H(\lambda_0) - \ln Z_\beta(\lambda_0)$, so that 
\begin{equation}
\label{eq:nonadiabaticity_first_term}
\begin{split}
    \Tr\{ &\rho(0)\ln\rho(0) \}\\ 
    &= \Tr\left\{ \sum_n\mu_n(\lambda_0)P_n(\lambda_0)\left[-\beta H(\lambda_0) - \ln Z_\beta(\lambda_0)\right] \right\}\\
    &= \sum_n\mu_n(\lambda_0)\Tr\{ P_n(\lambda_0)[-\beta E_n(\lambda_0) - \ln Z_\beta(\lambda_0)] \}\\
    &= \sum_n\Tr\{ P_n(\lambda_0) \}\mu_n(\lambda_0)\ln\mu_n(\lambda_0).
\end{split}
\end{equation}

Now we simplify the second term in the last line of Eq. (\ref{eq:nonadiabaticity_step1}). From $H(\lambda_0)=\sum_m E_m(\lambda_0)P_m(\lambda_0)$ and the adiabatic theorem $U_\text{ad}P_m(\lambda_0)U_\text{ad}^\dagger=P_m(\lambda_\tau)$, one concludes that 
\begin{equation}
\label{eq:nonadiabaticity_second_term}
\begin{split}
    \Tr\{ &U\rho(0)U^\dagger U_\text{ad}\ln\rho(0)U_\text{ad}^\dagger \}\\
    &= \Tr\{ U \rho(0)U^\dagger U_\text{ad}[-\beta H(\lambda_0) - \ln Z_\beta(\lambda_0)]U_\text{ad}^\dagger \}\\
    &= \sum_m\Tr\{ U \rho(0)U^\dagger [-\beta E_m(\lambda_0) - \ln Z_\beta(\lambda_0)]P_m(\lambda_\tau)\}\\
    &= \sum_{m,n}\Tr\{ UP_n(\lambda_0)U^\dagger P_m(\lambda_\tau) \}\mu_n(\lambda_0)\ln\mu_m(\lambda_0).
\end{split}
\end{equation}
Substitution of Eqs. (\ref{eq:nonadiabaticity_first_term}) and (\ref{eq:nonadiabaticity_second_term}) into Eq. (\ref{eq:nonadiabaticity_step1}) yields
\begin{equation}
\label{eq:aux D_ad=beta<X>}
\begin{split}
\mathcal{A} &= \sum_n\Tr\{ P_n(\lambda_0) \}\mu_n(\lambda_0)\ln\mu_n(\lambda_0)\\
&-\sum_{m,n}\Tr\{ UP_n(\lambda_0)U^\dagger P_m(\lambda_\tau) \}\mu_n(\lambda_0)\ln\mu_m(\lambda_0)
\end{split}
\end{equation}
Noting that $\Tr\{ P_n(\lambda_0) \} = \sum_m\Tr\{ UP_n(\lambda_0)U^\dagger P_m(\lambda_\tau) \}$, and using Eq. (\ref{eq:p(m,n)}), we rewrite Eq. (\ref{eq:aux D_ad=beta<X>}) as 
\begin{equation}
\label{eq:D_ad and X}
\begin{split}
\mathcal{A} &= \sum_{m,n}p(m,n)\ln\frac{\mu_n(\lambda_0)}{\mu_m(\lambda_0)}\\
&= \beta \sum_{m,n}p(m,n)E_{mn}(\lambda_0)= \beta\langle X\rangle,
\end{split}
\end{equation}
where we used Eq. (\ref{eq:<X>}) in the last line. This proves the relationship between $\mathcal{A}\equiv D[\rho(\tau)||\rho_\text{ad}(\lambda_\tau)]$ and the average of $X$.

We now prove that the excess work $W_\text{ex}$ is equal to $\langle Y \rangle$. Since the system is thermally isolated, the average work is given by the difference between the final and initial average energies, so that Eq. (\ref{eq:W_ex def1}) implies
\begin{equation}
\label{eq:W_ex def}
    \begin{split}
    W_\text{ex} &= \Tr\{ \rho(\tau)H(\lambda_\tau) - \rho(0)H(\lambda_0) \}\\
    & \quad\quad- \Tr\{ \rho_\text{ad}(\lambda_\tau)H(\lambda_\tau) - \rho(0)H(\lambda_0) \}\\
    &= \Tr\{ [\rho(\tau) - \rho_\text{ad}(\lambda_\tau)]H(\lambda_\tau)\}.
    \end{split}
\end{equation}
This can be rewritten as 
\begin{equation}
\label{eq:<Y>=W_ex}
    \begin{split}
    W_\text{ex} &= \sum_{m,n}\mu_n(\lambda_0)\Tr\{ [UP_n(\lambda_0)U^\dagger - P_n(\lambda_\tau)]\\
    & \ \ \ \ \ \ \ \ \ \ \ \ \ \ \ \ \ \ \ \ \ \ \ \ \ \  \times E_m(\lambda_\tau)P_m(\lambda_\tau) \}\\
    &= \sum_{m,n}\mu_n(\lambda_0)\Tr\{ UP_n(\lambda_0)U^\dagger P_m(\lambda_\tau)\}E_m(\lambda_\tau)\\
    & \ \ \ \ \ \ \ \ \ \ \ \ \ -\sum_n \mu_n(\lambda_0) \Tr\{ P_n(\lambda_\tau) \} E_n(\lambda_\tau)\\
    &= \sum_{m,n}\mu_n(\lambda_0)\Tr\{ UP_n(\lambda_0)U^\dagger P_m(\lambda_\tau)\}E_{mn}(\lambda_\tau)\\
    &= \sum_{m,n} p(m,n) E_{mn}(\lambda_\tau) = \langle Y\rangle, 
    \end{split}
\end{equation}
which proves that the average of $Y$ is equal to the excess work.
In the third equality, we used the hypothesis that the degree of degeneracy of each eigenspace is constant, $\Tr\{ P_n(\lambda_\tau) \} = \Tr\{ P_n(\lambda_0) \}$, as already stated in Eq.~(\ref{eq:degeneracy}).

Further insight can be gained by noting that the average of the quantity $Y'$ (defined in Eq. (\ref{eq:Y' def})) has a clear physical meaning in the specific case $\beta'=\overline\beta$, where $\overline \beta$ is an inverse temperature whose meaning and definition will be presented below.

In Thermodynamics, infinitely slow processes in thermally isolated systems conserve entropy and lead to an equilibrium \emph{macrostate}, which, according to Statistical Mechanics, should be described by a Gibbs state. The temperature of this state can then be defined as the temperature $1/k_B\overline\beta$ such that the thermodynamic entropy of the final Gibbs state is equal to that of the initial one,
\begin{equation}
    \label{eq:beta_bar def}
    \Tr\{ \Pi_{\overline\beta}(\lambda_\tau)\ln\Pi_{\overline\beta}(\lambda_\tau) \} \equiv \Tr\{ \Pi_\beta(\lambda_0)\ln\Pi_\beta(\lambda_0) \}.
\end{equation}
Thus, one can view $\overline\beta$ as the standard statistical-mechanical prediction for the temperature at the end of an infinitely slow process.

Now, let us define
\begin{equation}
    \overline Y \equiv \overline \beta Y + \overline \sigma,
\end{equation}
the value of $Y'$ for the specific case of $\beta'=\overline\beta$. Using Eq. (\ref{eq:<Y>=W_ex}), one sees that 
\begin{equation}
\label{eq:<Y_bar> = beta_bar W_ex + <sigma_bar>}
    \langle \overline Y \rangle = \overline \beta W_\text{ex} + \langle \overline \sigma \rangle.
\end{equation}
We must now rewrite $\langle \overline \sigma\rangle$. Starting from Eq. (\ref{eq:overline beta aux}) and using Eq. (\ref{eq:beta_bar def}), one can easily see that 
\begin{equation}
    \label{eq:<sigma_bar> = DeltaE_min}
    \langle \overline \sigma \rangle = \overline \beta\Tr\{ [\rho_\text{ad}(\lambda_\tau) - \Pi_{\overline \beta}(\lambda_\tau)]H(\lambda_\tau) \} 
\end{equation}
i.e., $\langle \overline \sigma\rangle$ is proportional to the difference between the average energies of the adiabatic state and the Gibbs state at temperature $\overline \beta$. Combining Eqs. (\ref{eq:W_ex def}), (\ref{eq:<Y_bar> = beta_bar W_ex + <sigma_bar>}), and (\ref{eq:<sigma_bar> = DeltaE_min}) yields
\begin{equation}
    \langle \overline Y \rangle = \overline \beta W - \overline\beta\Tr\{ \Pi_{\overline\beta}(\lambda_\tau)H(\lambda_\tau) - \Pi_\beta(\lambda_0)H(\lambda_0) \}.
\end{equation}
Note that the second term is just the work that one would calculate for an isentropic quasistatic process using standard Statistical Mechanics. We call it $W_\text{ise}$,
\begin{equation}
    W_\text{ise} \equiv \Tr\{ \Pi_{\overline\beta}(\lambda_\tau)H(\lambda_\tau) - \Pi_\beta(\lambda_0)H(\lambda_0) \}.
\end{equation}
Then, we can write
\begin{equation}
\label{eq:<Y_bar> = beta_bar W_th}
    \langle \overline Y\rangle = \overline \beta [W - W_\text{ise}]\equiv \overline \beta W_\text{ex}^\text{th},
\end{equation}
where we defined a quantity we call the ``thermodynamic excess work'', as it represents the additional work done in comparison to the thermodynamic-statistical-mechanical prediction $W_\text{ise}$ for an isentropic process.\footnote{By naming $W-W_\text{ise}$ the \textit{thermodynamic} excess work, we intend to highlight the thermodynamic motivation (further explained in Sec. \ref{sec:Y and minimal work}) for defining this quantity. This name, however, should not be taken as a suggestion that the excess work $W_\text{ex} = W - W_\text{ad}$ is not a thermodynamic quantity.}. The equation above clarifies the physical meaning of $\overline Y$.

\section{Connection to irreversibility and the Second Law}
\label{sec:irreversibility}

We mentioned in Sec.~\ref{sec:intro} that fluctuation theorems and also the quantities involved in them, are usually connected to statements of the Second Law and, therefore, to some notion of irreversibility. In this section, we make this connection explicit by relating the averages of $X$ and $Y$ to two formulations of the Second Law, namely Thomson's formulation and the minimal work principle.

\subsection{Irreversibility and $\langle X\rangle$}
\label{sec:X and irr}

In this section, we develop several arguments that clarify the connection between the nonadiabaticity parameter and the notion of irreversibility. We begin by discussing Thomson's formulation of the Second Law.

Thomson's formulation was proved rigorously under the assumptions of Hamiltonian dynamics and an initial Gibbs state\footnote{In fact, the initial state can be any passive state, not necessarily a Gibbs state. In this work, however, we consider only the Gibbs state as the initial state.} with positive temperature \cite{AllahverdyanNieuwenhuizen2002_SecondLawTheorem}. It states that there can be no work extraction from a cyclic variation of an external parameter, provided that the system is initially in a Gibbs state. In other words, the average work in a cyclic process starting from equilibrium is non-negative. 

In Thomson's formulation, the value of the average work done during the cyclic process distinguishes between reversible and irreversible processes: if it is zero, the process is reversible; if it is positive, the process is irreversible. The greater the work, the more irreversible the process. Although this reasoning only applies to cyclic processes, we may devise a way to extend it to non-cyclic ones. Imagine that after a given non-cyclic protocol, we supplement the process with an adiabatic protocol that goes through the same values of the external parameter, but backwards and infinitely slowly\footnote{It is important to note that, just before this backward adiabatic process, we do not need to perform any measurement. We also do not time-reverse the final state of the original process. We simply perform the adiabatic backward protocol on the final state of the original process.}. We thus construct a cyclic process from the original non-cyclic one. Since the backward protocol is infinitely slow, we can imagine that it does not introduce any further ``irreversibility'' to the original non-cyclic process, that is, all the irreversibility in the cyclic process is due to the original finite-time protocol. The average work,
\begin{equation}
    W_c \equiv W + W_\text{ad}^\text{back}\,,
\end{equation}
performed during the imaginary cyclic process, which (according to the reasoning at the beginning of this paragraph) is a measure of its irreversibility, can then also be viewed as a measure of the irreversibility of the original non-cyclic process.

\subsubsection{Nonadiabaticity and work absorption}
\label{sec:work absorption}

From the definitions of $W_c$ and $w_c$, and from the fact that $w_c=X$ (see Sec.~\ref{sec:interpretation} and app.~\ref{app:X=w_c}), we have $W_c = \langle w_c\rangle =\langle X\rangle$. Therefore, the following equation holds\footnote{One should be careful to avoid confusion regarding the quantities involved in Eq. (\ref{eq:W_c = D_ad}). $W$, $\rho(\tau)$, and $\rho_\text{ad}(\lambda_\tau)$ all refer to quantities and density operators evaluated at the end of the original non-cyclic process. Only $W_\text{ad}^\text{back}$ refers to the imaginary backward protocol.}:
\begin{equation}
    \label{eq:W_c = D_ad}
    W_c \equiv W + W_\text{ad}^\text{back} =  \beta^{-1} D[\rho(\tau)||\rho_\text{ad}(\lambda_\tau)],
\end{equation}
where we used Eq. (\ref{eq:D_ad and X}). This means that the work absorbed by the system under the imaginary cycle is proportional to the nonadiabaticity parameter. Combining the equation above with the reasoning from the last paragraph of Sec.~\ref{sec:X and irr}, we conclude that the nonadiabaticity parameter (and thus also $X$) can be viewed as a measure of irreversibility (in the sense of Thomson's formulation) of a given not necessarily cyclic process, as it is proportional to $W_c$. The strong physical significance of this interpretation establishes Eq. (\ref{eq:W_c = D_ad}) as one of the central results of this work.

Let us discuss a subtlety regarding the definition of $W_\text{ad}^\text{back}$. Because we have constructed $W_c$ without intermediate measurements during the imaginary cycle, one could argue (in view of the two-point measurement scheme) that $W_\text{ad}^\text{back}$ is a meaningless quantity, as it is not experimentally accessible (energy measurements are required at the beginning and the end of a given process to obtain the work done during it). Nevertheless, this difficulty can be surpassed in two different ways. The first is to note that, regarding the contents of Eq.~(\ref{eq:W_c = D_ad}), the splitting of $W_{c}$ as $W + W_\text{ad}^\text{back}$ is not necessary since the physically relevant (and experimentally accessible) quantity is $W_c$ itself. The second way to overcome the posted criticism is to note that the value of $W_\text{ad}^\text{back}$ as defined by not performing an intermediate measurement after the original forward process is actually the same as if it were defined \textit{with} the intermediate measurement. This can be viewed as a consequence of the fact that $X=w_c$. Therefore, even if we define $W_\text{ad}^\text{back}$ without the intermediate measurement, it is still measurable due to the equivalence with the intermediate measurement case.

Interestingly, the form of Eq. (\ref{eq:W_c = D_ad}) is very similar to that of the well-known equation \cite{VaikuntanathanJarzynski2009_DissipationLag, EspositoVandenBroeck2011_SecondLawLandauer}
\begin{equation}
    \label{eq:W_diss = D[rho||Pi]}
    W_\text{diss}\equiv W - \Delta F = \beta^{-1}D[\rho(\tau)||\Pi_\beta(\lambda_\tau)],
\end{equation}
where $W_\text{diss}\equiv W - \Delta F$ is the so-called dissipated work. The only caveat is that $W_\text{ad}^\text{back}$ depends on the final state of the original process and thus on the forward protocol, whilst $\Delta F$ only depends on the endpoints $\lambda_0$ and $\lambda_\tau$.

\subsubsection{Nonadiabaticity and time-reversal asymmetry}
\label{sec:time-reversal asymmetry}

It is known that the dissipated work $W_\text{diss}$ also satisfies the relation \cite{KawaiParrondoVanDenBroeck2007_DissipationPhaseSpace, ParrondoVanDenBroeckKawai2009_EntropyArrowTime, RubinoManzanoRozemaWaltherParrondoBrukner2022_InferringWork}
\begin{equation}
    \label{eq:W_diss = D_rev}
    \beta W_\text{diss} = D[\rho(t)||\Theta^\dagger\tilde\rho(\tau-t)\Theta], \forall t \in [0,\tau].
\end{equation}
In the equation above, $\tilde\rho(\tau - t) = \tilde U(\tau - t, 0) \tilde\rho(0) \tilde U^\dagger(\tau - t, 0) = \tilde U(\tau - t, 0)\Theta \Pi_\beta(\lambda_\tau)\Theta^\dagger \tilde U^\dagger(\tau - t, 0)$ is the state at time $\tau - t$ in the time-reversed process, which for this equation starts at $\tilde \rho(0)  =\Theta \Pi_\beta(\lambda_\tau)\Theta^\dagger$ instead of $\Theta\rho_\text{ad}(\lambda_\tau)\Theta^\dagger$. This equation relates $W_\text{diss}$ to a measure of time-reversal asymmetry, i.e., $D[\rho(t)||\Theta^\dagger\tilde\rho(\tau-t)\Theta]$, which compares the state at an arbitrary\footnote{Interestingly, although $W_\text{diss}$ is the dissipated work regarding the entire process, the relative entropy in the right-hand side of Eq. (\ref{eq:W_diss = D_rev}) can be evaluated at any time $t\in [0,\tau]$.} time $t$ in the forward process and the state at the corresponding time $\tau - t$ in the time-reversed one. Given the similarities between Eqs. (\ref{eq:W_c = D_ad}) and (\ref{eq:W_diss = D[rho||Pi]}), one may inquire whether $W_c = \beta^{-1} D[\rho(\tau)||\rho_\text{ad}(\lambda_\tau)]$ satisfies an equation similar to Eq. (\ref{eq:W_diss = D_rev}). Indeed, it is straightforward to show that
\begin{equation}
    \label{eq:W_c = D_rev}
    \beta W_c = D[\rho(t)||\Theta^\dagger\tilde\rho(\tau-t)\Theta], \forall t \in [0,\tau],
\end{equation}
where now the initial distribution of the time-reversed process is $\tilde\rho(0)=\Theta\rho_\text{ad}(\lambda_\tau)\Theta^\dagger$, the same as for the DFT for $X$. Thus, in Eq. (\ref{eq:W_c = D_rev}), $\tilde\rho(\tau-t)$ is given by
\begin{equation}
\begin{split}
\tilde\rho(\tau-t) &= \tilde U(\tau-t,0)\Theta \rho_\text{ad}(\lambda_\tau)\Theta^\dagger\tilde U^\dagger(\tau-t, 0)\\
&= \Theta U^\dagger(\tau,t)\rho_\text{ad}(\lambda_\tau)U(\tau,t)\Theta^\dagger,
\end{split}
\end{equation}
where we used $\tilde U(\tau-t,0)=\Theta U^\dagger(\tau,t)\Theta^\dagger$ and the fact that $\Theta^\dagger\Theta=\Theta\Theta^\dagger = \mathbb{1}$. To prove Eq. (\ref{eq:W_c = D_rev}), one then simply notes that
\begin{equation}
\label{eq:aux W_c = D_rev}
\begin{split}
    \Tr\{ \rho(t)&\ln[\Theta^\dagger\tilde\rho(\tau-t)\Theta] \}\\
    &= \Tr\{ \rho(t)\ln[U^\dagger(\tau,t)\rho_\text{ad}(\lambda_\tau)U(\tau,t)] \}\\
    &= \Tr\{ U(\tau,t)\rho(t)U^\dagger(\tau,t)\ln\rho_\text{ad}(\lambda_\tau) \}\\
    &= \Tr\{ \rho(\tau)\ln\rho_\text{ad}(\lambda_\tau) \}.
\end{split}
\end{equation}

Hence, 
\begin{equation}
\begin{split}
    D[\rho&(t)||\Theta^\dagger\tilde\rho(\tau-t)\Theta]\\
    &= \Tr\{ \rho(t)\ln\rho(t) \} -  \Tr\{ \rho(t)\ln[\Theta^\dagger\tilde\rho(\tau-t)\Theta] \}\\
    &= \Tr\{ \rho(\tau)\ln\rho(\tau) \} - \Tr\{ \rho(\tau)\ln\rho_\text{ad}(\lambda_\tau) \}\\
    &= D[\rho(\tau)||\rho_\text{ad}(\lambda_\tau)] = \beta W_c,
\end{split}
\end{equation}
which proves Eq. (\ref{eq:W_c = D_rev}). In the third line of the equation above, we used the invariance of the von-Neumann entropy under unitary evolution and Eq. (\ref{eq:aux W_c = D_rev}). In the fourth line, Eq. (\ref{eq:W_c = D_ad}) was used.

Equation (\ref{eq:W_c = D_rev}) implies that, just as $W_\text{diss}$, the quantity $W_c$ is also related to a measure of time-reversal asymmetry or distinguishability between the forward and time-reversed processes. For Eq. (\ref{eq:W_c = D_rev}), however, the initial state of the time-reversed process is $\Theta\rho_\text{ad}(\lambda_\tau)\Theta^\dagger$. Once again, the reason for this can be better interpreted through the picture of ``imaginary cycles'' introduced in Sec.~\ref{sec:interpretation}; see also Fig. \ref{fig:X_DFT}. When viewing the problem through this lens, one sees that the right-hand side of Eq. (\ref{eq:W_c = D_rev}) measures the distinguishability between the states in the ``original'' portion of the forward and time-reversed cycles, the latter cycle being obtained simply by time reversing the protocol and initial state $\Pi_\beta(\lambda_0)$ of the former.

This discussion highlights that the quantity $W_c$, which is related to irreversibility according to the argument presented in Sec.~ \ref{sec:X and irr}, is equal to a measure of the time-reversal asymmetry between the imaginary forward cycle and the imaginary time-reversed cycle.

\subsubsection{Nonadiabaticity, coherence and transitions}
\label{sec:coherence and transitions}

Only in this section, we assume the projectors $P_n(\lambda)$ satisfy $\Tr\{P_n(\lambda_t)\} = 1, \forall t \in [0, \tau]$, i.e., the spectrum is non-degenerate at all times. 
In this context, Ref. \cite{FrancicaGooldPlastina2019_CoherenceThermo} clarifies the role of coherence and transitions in entropy production by showing that the dissipated work $W_\text{diss}=\beta^{-1}D[\rho(\tau)||\Pi_\beta(\lambda_\tau)]$ can be decomposed into a sum of their contributions. Remarkably, the authors show that the nonadiabaticity parameter $\mathcal{A}$ can also be expressed as a sum of these contributions: both $W_\text{diss}$ and $\mathcal{A}$ can be written in the form
\begin{equation}
    \label{eq:D = coherence + transition}
    \mathcal{D} = D[\rho(\tau)||\Delta \rho(\tau)] + D[\Delta \rho(\tau)||\mathcal{\xi}],
\end{equation}
where $\mathcal{D}$ represents either $\beta W_\text{diss} = D[\rho(\tau)||\Pi_\beta(\lambda_\tau)]$ or $\mathcal{A}=D[\rho(\tau)||\rho_\text{ad}(\lambda_\tau)]$. For $\mathcal{D}=\beta W_\text{diss}$, $\xi$ corresponds to the Gibbs state $\Pi_\beta(\lambda_\tau)$, whilst for $\mathcal{D}=\mathcal{A}$ it is the adiabatic state, $\xi=\rho_\text{ad}(\lambda_\tau)$. In both cases, $\Delta \rho(\tau)$ represents the operator obtained by removing all the coherences (with respect to the final instantaneous energy eigenbasis) from $\rho(\tau)$, that is
\begin{equation}
    \Delta\rho(\tau) = \sum_n P_n(\lambda_\tau)\rho(\tau)P_n(\lambda_\tau).
\end{equation}
Thus, one may interpret the first term of the sum in Eq. (\ref{eq:D = coherence + transition}) as a measure of the coherence generated by the driving, as it measures how different the state $\rho(\tau)$ (which might have coherences) is from the completely incoherent $\Delta \rho(\tau)$. On the other hand, the second term can be seen as a quantifier of ``unwanted'' transitions, as it measures the mismatch between the populations of $\rho(\tau)$ and those of $\xi$. For $W_\text{diss}$, these transitions are measured in relation to $\Pi_\beta(\lambda_\tau)$, whilst for $\mathcal{A}$, they are measured in relation to the adiabatic state. This hints at the idea that $W_\text{diss}$ is an appropriate irreversibility quantifier only when the system's final state is $\Pi_\beta(\lambda_\tau)$, which occurs if, at the end of the process, the system is allowed to equilibrate with a weakly coupled ideal heat bath at temperature $T=1/k_B\beta$ (this point is discussed in more detail in App.~\ref{app:W_diss and entropy production}). On the other hand, the term that measures transitions with respect to the adiabatic state (which is the state obtained under ``transitionless'' evolution) is physically relevant in cases where there is no thermal contact at the end of the protocol; thus, one would expect the nonadiabaticity parameter to be an appropriate quantifier of irreversibility in such cases. This topic is further discussed in the next subsection.

Lastly, we should mention that, as noted in Ref. \cite{FrancicaGooldPlastina2019_CoherenceThermo}, the coherence term $D[\rho(\tau)||\Delta\rho(\tau)]$ is simply the change in diagonal entropy \cite{Polkovnikov2011_DiagonalEntropy}. Besides, for slow processes, the leading contribution to the nonadiabaticity parameter is precisely the change in diagonal entropy, as shown in \cite{FrancicaGooldPlastina2019_CoherenceThermo}.

\subsubsection{The nonadiabaticity parameter as a quantifier of irreversibility}

In the last few subsections, we have been exploring the comparison between the dissipated work and the nonadiabaticity parameter. The reason for this is that the dissipated work is a well-established quantifier of irreversibility that yields the thermodynamic entropy production under certain conditions, among which is the requirement that the final state of the system is an equilibrium state with the same temperature as the initial state (we expand on these conditions in App.~\ref{app:W_diss and entropy production}). Evidently, this condition is generally not satisfied for systems that evolve in thermal isolation and are not allowed to equilibrate with a heat bath by the end of the protocol. Therefore, in this scenario, the dissipated work cannot be related to irreversibility in any direct way. In this subsection, we argue that, instead, the nonadiabaticity parameter is an interesting quantifier of irreversibility in this case. We do so by enumerating the facts that support our claim, most of which were already discussed to some extent in the previous pages:

\begin{enumerate}
    \item The nonadiabaticity parameter $\mathcal{A}=\beta\langle X\rangle$ is the average of a quantity that satisfies both an integral and a detailed fluctuation theorem, as shown in Sec.~\ref{sec:X_DFT}. Since fluctuation theorems are strongly related to the idea of irreversibility, so are the quantities that satisfy them;
    \item The nonadiabaticity parameter is proportional to the work absorbed by the system during the imaginary cycle, a quantity we argued quantifies irreversibility (see Sec.~\ref{sec:X and irr});
    \item $\mathcal{A}$ is equal to a measure of the time-reversal asymmetry of the imaginary cycles, as shown in Sec.~\ref{sec:time-reversal asymmetry};
    \item Without appealing to any particular definition of entropy, one may define reversibility in the following way. A reversible process is such that, after it is carried out, there exists a supplementary process\footnote{Of course, this supplementary process must follow the laws of Physics. For example, for a thermally isolated system, we cannot choose a supplementary process represented by an anti-unitary evolution operator, as the evolution is dictated by Hamiltonian dynamics.} that returns all physically relevant observables of the system and its surroundings to their initial values. In the case of thermally isolated driven quantum systems, this amounts to recovering the initial values of both the external parameter and the system's observable averages. Under this definition, one can easily see that adiabatic processes in thermally isolated systems starting from Gibbs states are reversible: after a given adiabatic process, one may simply perform another adiabatic process taking the external parameter back to its initial value and this restores the system's initial density matrix, and thus all its observable averages;
    \item A good quantifier of irreversibility should be able to capture the reversible case by yielding zero for adiabatic processes. Indeed, by its very definition, $\mathcal{A}$ satisfies this requirement;
    \item The nonadiabaticity parameter measures the distinguishability between the actual state $\rho(\tau)$ of the system and the state $\rho_\text{ad}(\lambda_\tau)$ that would be achieved under a reversible process, i.e., it measures the breaking of reversibility; 
    \item Just as $W_\text{diss}$, the nonadiabaticity parameter can be decomposed into a term due to coherence generation and a term due to population mismatch (see Sec.~\ref{sec:coherence and transitions}). However, the population mismatch for $W_\text{diss}$ is measured against a state that is generally never achieved by the unitary dynamics, while for $\mathcal{A}$ it is measured against the state achieved under adiabatic (``transitionless''), reversible evolution. This makes $\mathcal{A}$ more appropriate for the case of thermally isolated driven systems that are not allowed to equilibrate with a heat bath at the end of the protocol;
    \item For slow processes, the leading contribution to $\mathcal{A}$ is the change in diagonal entropy \cite{FrancicaGooldPlastina2019_CoherenceThermo}, which has been shown under fairly general conditions to exhibit some properties of the entropy production from standard Thermodynamics \cite{Polkovnikov2011_DiagonalEntropy}. However, the change in diagonal entropy corresponds only to the coherence term $D[\rho(\tau)||\Delta\rho(\tau)]$, thus ignoring the contributions of ``unwanted'' transitions, which become relevant for faster processes. On the other hand, the nonadiabaticity parameter takes both contributions into account. 

\end{enumerate}

Given these points, it seems reasonable to conclude that the nonadiabaticity parameter $\mathcal{A}$ is an interesting quantity that warrants further investigation, potentially as a quantifier of irreversibility.

\subsection{The minimal work principle and $\langle Y \rangle$ }
\label{sec:Y and minimal work}

In this section, we explore the connection between $\langle Y \rangle=W_\text{ex}$ and the so-called minimal work principle.

The minimal work principle is a formulation of the Second Law that states the following: the work done on a thermally isolated driven system initially in equilibrium is minimal for the slowest realization of the process \cite{AllahverdyanNieuwenhuizen2005_MinimalWorkPrinciple}, i.e., $W_\text{ex}\ge 0$. In this formulation, adiabatic evolution yields $W_\text{ex} = 0$, indicating that the process is reversible, while $W_\text{ex}>0$ indicates an irreversible process. Then, one can view the quantity $\langle Y \rangle=W_\text{ex}$ as a quantifier of irreversibility according to this formulation. Furthermore, the minimal work principle was shown to be equivalent to the law of entropy increase for volume entropy \cite{Campisi2008_VolumeEntropy, Tasaki2000}.

The minimal work principle was rigorously proven in Ref. \cite{AllahverdyanNieuwenhuizen2005_MinimalWorkPrinciple} under the following assumptions: the initial state is passive, the evolution is governed by Hamiltonian dynamics, and there are no level crossings. 

However, there is a subtle difference between the statement of the minimal work principle given above and the corresponding standard thermodynamic statement, which is implied by the law of entropy increase. We briefly present the derivation of the thermodynamic statement before comparing it to the minimal work principle from \cite{AllahverdyanNieuwenhuizen2005_MinimalWorkPrinciple}, $W_\text{ex}\ge0$. 
The derivation is based exclusively on thermodynamic considerations and we follow closely the reasoning in the Appendix of Ref.~\cite{Jarzynski2020}. Let the system start at an equilibrium state with external parameter $\lambda_0$ and thermodynamic entropy $S$, denoted by $(\lambda_0, S)$. A finite-time protocol $\lambda_t$ is carried out, driving the system out of equilibrium. After the end of the protocol, the system self-equilibrates\footnote{As mentioned, this derivation is based exclusively on thermodynamic reasoning. The actual occurrence (or lack thereof) of self-equilibration and the systems for which it occurs are topics of intense research.} (no thermal contact is assumed), reaching the equilibrium state $(\lambda_\tau, S')$. Then, the work done on the system is given by
\begin{equation}
    W = E(\lambda_\tau,S') - E(\lambda_0,S),
\end{equation}
where, in general, $S'\ge S$. Since $\partial E/\partial S = T \ge 0$, we have $E(\lambda_\tau,S')\ge E(\lambda_\tau,S)$, which implies
\begin{equation}
    W \ge E(\lambda_\tau,S) - E(\lambda_0,S).
\end{equation}
Since equilibrium macrostates can be described by Gibbs states according to standard Statistical Mechanics, the difference $E(\lambda_0,S) - E(\lambda_\tau,S)$ can be expressed as 
\begin{equation}
\begin{split}
    E(\lambda&_\tau,S) - E(\lambda_0,S)\\
    &= \Tr\{\Pi_{\overline\beta}(\lambda_\tau)H(\lambda_\tau)  - \Pi_\beta(\lambda_0)H(\lambda_0)\} = W_\text{ise}.
\end{split}
\end{equation}
Thus, the minimal work principle as implied by standard thermodynamic considerations (we shall refer to it as ``thermodynamic minimal work principle'' from now on) is just 
\begin{equation}
\label{eq:thermodynamic_minimal_work_principle}
    W - W_\text{ise} \equiv W_\text{ex}^\text{th} \ge 0.
\end{equation}

One can see that the reference to which the work is compared is different across these principles, $W_\text{ad}\neq W_\text{ise}$. This is a direct consequence of the fact that the adiabatic state $\rho_\text{ad}(\lambda_\tau)$ is different from the isentropic Gibbs state $\Pi_{\overline\beta}(\lambda_\tau)$. Indeed, one can easily see from the definitions of $W_\text{ad}$ and $W_\text{ise}$ that
\begin{equation}
\label{eq:W_ad-W_ise}
\begin{split}
    \overline\beta (W_\text{ad} - W_\text{ise}) &= \overline\beta\Tr\{ [\rho_\text{ad}(\lambda_\tau) - \Pi_{\overline\beta}(\lambda_\tau)]H(\lambda_\tau) \}\\
    &= -\Tr\{ [\rho_\text{ad}(\lambda_\tau) - \Pi_{\overline\beta}(\lambda_\tau)]\ln\Pi_{\overline\beta}(\lambda_\tau) \}\\
    &= \Tr\{\rho_\text{ad}(\lambda_\tau)[\ln\rho_\text{ad}(\lambda_\tau) - \ln\Pi_{\overline\beta}(\lambda_\tau)]\}\\
    &= D[\rho_\text{ad}(\lambda_\tau)||\Pi_{\overline\beta}(\lambda_\tau)] \ge0.
\end{split}
\end{equation}
In the third line, we used the fact that, by definition, $\Tr\{\rho_\text{ad}(\lambda_\tau)\ln\rho_\text{ad}(\lambda_\tau)\} = \Tr\{\Pi_{\overline\beta}(\lambda_\tau)\ln\Pi_{\overline\beta}(\lambda_\tau)\}$. Equation (\ref{eq:W_ad-W_ise}) shows that the difference between $W_\text{ad}$ and $W_\text{ise}$ is proportional to a measure (the relative entropy) of how different the adiabatic state is from the isentropic Gibbs state.

As a side note, we also show a similar relation between $W_\text{ise}$ and $\Delta F$. We use the fact $\Delta F = \Delta E - \beta^{-1}\Delta S$, where $\Delta E = \Tr\{\Pi_\beta(\lambda_\tau)H(\lambda_\tau) -\Pi_\beta(\lambda_0)H(\lambda_0)\}$ and $\Delta S = -\Tr\{ \Pi_\beta(\lambda_\tau)\ln\Pi_\beta(\lambda_\tau) \} + \Tr\{ \Pi_\beta(\lambda_0)\ln\Pi_\beta(\lambda_0) \}$. Then, one can show that:
\begin{equation}
    \begin{split}
    \beta(W_\text{ise}& - \Delta F) = \beta\Tr\{ [\Pi_{\overline\beta}(\lambda_\tau) - \Pi_{\beta}(\lambda_\tau)]H(\lambda_\tau) \} + \Delta S\\
    &= \Delta S -\Tr\{[\Pi_{\overline\beta}(\lambda_\tau) - \Pi_{\beta}(\lambda_\tau)]\ln\Pi_{\beta}(\lambda_\tau)\}\\
    &= \Tr\{\Pi_\beta(\lambda_0)\ln\Pi_\beta(\lambda_0)\} - \Tr\{ \Pi_{\overline\beta}(\lambda_\tau) \ln\Pi_\beta(\lambda_\tau) \}\\
    &= \Tr\{ \Pi_{\overline\beta}(\lambda_\tau)\ln\Pi_{\overline\beta}(\lambda_\tau) \} - \Tr\{ \Pi_{\overline\beta}(\lambda_\tau) \ln\Pi_\beta(\lambda_\tau) \}\\
    &= D[\Pi_{\overline\beta}(\lambda_\tau)||\Pi_\beta(\lambda_\tau)] \ge 0.
    \end{split}
\end{equation}
Therefore, provided that $\beta,\overline\beta\ge 0$, the following inequalities hold:
\begin{equation}
\label{eq:inequality excess work}
    \begin{split}
    &W_\text{ad}\ge W_\text{ise}\ge\Delta F\\
    \text{or} \quad &W_\text{ex}\le W_\text{ex}^\text{th}\le W_\text{diss}.
    \end{split}
\end{equation}
Hence, among $W_\text{ad}$, $W_\text{ise}$, and $\Delta F$, the former is the sharpest lower bound on the work done on a system with no level crossings.

Now that the distinction between the minimal work principles $W_\text{ex}\ge0$ and $W_\text{ex}^\text{th}\ge0$ has been made clear, we may ask whether any of them can be obtained via the fluctuation theorems derived in this paper. This was the original objective in \cite{Jarzynski2020}: to derive, via fluctuation relations, an inequality stronger than $W-\Delta F\ge0$, which can itself be obtained via the Jarzynski equality $\langle e^{-\beta(w-\Delta F)}\rangle=1$. 

It is straightforward to see that the thermodynamic minimal work principle is indeed attainable via fluctuation theorems. It suffices to take the IFT in Eq. (\ref{eq:Y IFT}) for the special case $\beta'=\overline\beta$, apply Jensen's inequality to get $\langle \overline Y\rangle \ge 0$, and use Eq. (\ref{eq:<Y_bar> = beta_bar W_th}) to conclude that:
\begin{equation}
\label{eq:W_th>0 from IFT}
    \langle e^{-\overline Y}\rangle = 1 \Longrightarrow W_\text{ex}^\text{th}\ge0, \text{ provided that } \overline \beta \ge 0.
\end{equation}

Now, one may ask whether $W_\text{ex}$ is also attainable via Eq. (\ref{eq:Y IFT}). To investigate this, we search for the best bound on the excess work among all possible bounds implied by Eq. (\ref{eq:Y IFT}). Applying Jensen's inequality to Eq.~(\ref{eq:Y IFT}) yields
\begin{equation}
\begin{split}
    \beta'&W_\text{ex} = \beta'\langle Y\rangle \ge -\langle \sigma'\rangle,\\
    \text{or}\ \ \ \ \ \ \  &W_\text{ex}\ge-\langle \sigma' \rangle/\beta' \text{ if } \beta'\ge0
\end{split}
\end{equation}
Restricting ourselves to non-negative $\beta'$, we now search for the $\beta'$ that gives the sharpest lower bound on the excess work. To do so, we search for the maximum points of (see Eq. (\ref{eq:overline beta aux}))
\begin{equation}
\label{eq:-<sigma'>/beta'}
-\frac{\langle \sigma'\rangle}{\beta'} = -\langle H(\lambda_\tau)\rangle_\text{ad} - \frac{\ln Z_{\beta'}(\lambda_\tau)}{\beta'} + \frac{S_\beta( \lambda_0)}{\beta'},
\end{equation}
where we defined $S_\beta( \lambda)\equiv -\Tr\{\Pi_\beta(\lambda) \ln\Pi_\beta(\lambda) \}$ and $\langle H(\lambda_\tau)\rangle_\text{ad}\equiv \Tr\{ \rho_\text{ad}(\lambda_\tau)H(\lambda_\tau) \}$ in order to shorten the notation. Differentiating with respect to $\beta'$ yields
\begin{equation}
\label{eq:first_derivative}
    \begin{split}
    -\frac{\partial}{\partial \beta'} \frac{\langle \sigma' \rangle}{\beta'} &= \frac{\ln Z_{\beta'}(\lambda_\tau)}{\beta'^2} - \frac{1}{\beta'} \frac{\partial}{\partial \beta'}\ln Z_{\beta'}(\lambda_\tau) - \frac{S_\beta(\lambda_0)}{\beta'^2}\\
    &= \frac{1}{\beta'^2}[\beta'\langle H(\lambda_\tau)\rangle' + \ln Z_{\beta'}(\lambda_\tau)] - \frac{S_\beta(\lambda_0)}{\beta'^2}\\
    &= \frac{S_{\beta'}(\lambda_\tau)}{\beta'^2} - \frac{S_\beta(\lambda_0)}{\beta'^2},
    \end{split}
\end{equation}
where $\langle H(\lambda_\tau)\rangle'\equiv \Tr\{ \Pi_{\beta'}(\lambda_\tau)H(\lambda_\tau) \}$. We see that $\beta'=\overline \beta$ is a stationary point. We differentiate again to check whether it is a point of maximum or minimum. 
\begin{equation}
\label{eq:second derivative}
\begin{split}
    -\frac{\partial^2}{\partial \beta'^2} \frac{\langle \sigma' \rangle}{\beta'}  &= \frac{\partial}{\partial \beta'}\left\{\frac{1}{\beta'^2}[S_{\beta'} (\lambda_\tau) - S_\beta(\lambda_0)] \right\}\\
    &= -\frac{2}{\beta'^3}[S_{\beta'}(\lambda_\tau) - S_\beta(\lambda_0)] + \frac{1}{\beta'^2}\frac{\partial}{\partial \beta'}S_{\beta'}(\lambda_\tau)\\
    &\le -\frac{2}{\beta'^3}[S_{\beta'}(\lambda_\tau) - S_\beta(\lambda_0)]
\end{split}
\end{equation}
In the third line, we used the fact that $\frac{\partial}{\partial\beta'}S_{\beta'}(\lambda_\tau)\le0, \forall \beta'\ge0$. Equation (\ref{eq:second derivative}) evaluated at $\beta'=\overline\beta$ shows that $\overline \beta$ is a maximum point of $-\langle\sigma'\rangle/\beta'$. We show that it is a global maximum in App.~\ref{app:global maximum}. Thus, the sharpest lower bound on the excess work that can be provided by the IFT is 
\begin{equation}
\label{eq:lower bound}
    W_\text{ex} \ge -\frac{\langle \overline \sigma\rangle }{\overline \beta}= -\frac{D[\rho_\text{ad}(\lambda_\tau)||\Pi_{\overline\beta}(\lambda_\tau)]}{\overline\beta},
\end{equation}
where we used Eq. (\ref{eq:<sigma_bar> = D}) to write the equality. This is, however, just Eq. (\ref{eq:W_th>0 from IFT}). Hence, the IFT for $Y$ does not imply the minimal work principle $W_\text{ex}\ge0$; the sharpest bound that follows from it is the thermodynamic minimal work principle $W_\text{ex}^\text{th}\ge0$.

Because $W_\text{ex}^\text{th}$ is non-negative and is the average of a quantity $\overline Y$ that satisfies fluctuation theorems, one could be tempted to view it as an interesting quantifier of irreversibility for thermally isolated driven systems. However, it generally does not approach zero in the adiabatic limit\footnote{One can see this by noting that $W_\text{ex}^\text{th} = W_\text{ex} + \langle\overline\sigma\rangle/\overline\beta$ and that $W_\text{ex}$ is, by definition, zero for adiabatic processes, while $\langle\overline\sigma\rangle = D[\rho_\text{ad}(\lambda_\tau)||\Pi_{\overline\beta}(\lambda_\tau)]$ is not necessarily zero.}, thus failing to capture the fact that adiabatic processes are reversible. Nevertheless, for large systems obeying the standard assumptions of Thermodynamics, one could expect the distinguishability between $\rho_\text{ad}(\lambda_\tau)$ and $\Pi_{\overline\beta}(\lambda_\tau)$ to be negligible, so that $W_\text{ex}^\text{th}$ and $W_\text{ex}$ would become equivalent for all practical purposes. Of course, this claim requires further investigation to be made rigorous.

\section{Examples}
\label{sec:examples}

In this section, we illustrate our results via two examples. First, we consider a non-integrable Ising chain. Then, we discuss an important class of systems whose parametric change of the energy spectrum follows a specific rule.

\subsection{Non-integrable Ising chain}
\label{sec:nonintegrable ising}

\begin{figure}[b]
    \centering
    \includegraphics[width=0.90\linewidth]{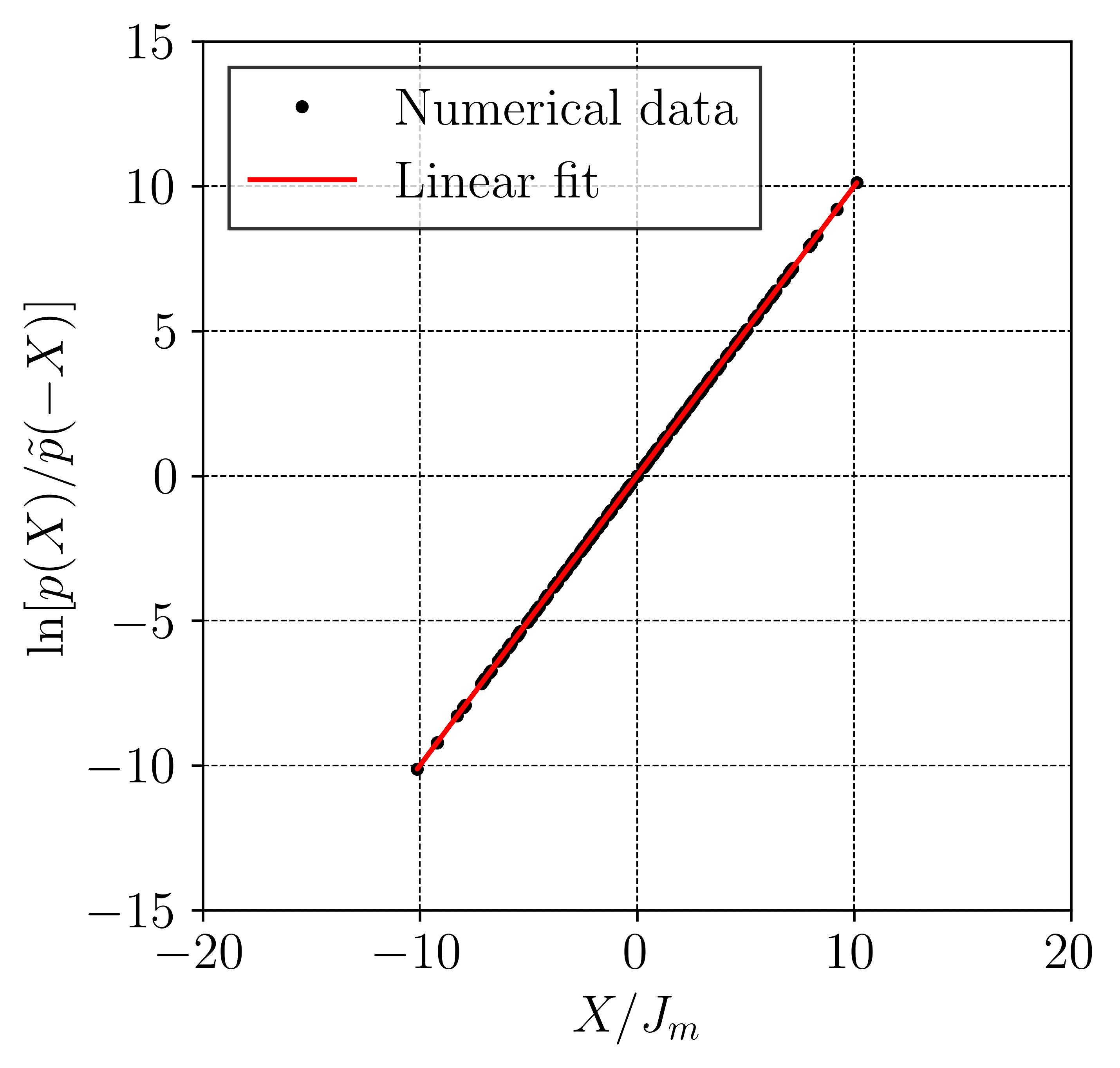}
    \caption{Numerical verification of the DFT for $X$, Eq. (\ref{eq:DFT for X}). The calculations were performed on the 4-spin Ising chain with the Hamiltonian of Eq. (\ref{eq:nonintetgrable_open}), protocol of Eq. (\ref{eq:lambda_protocol ising}), initial inverse temperature $\beta=J_m^{-1}$ and process duration $\tau=\hbar/J_m$.}
    \label{fig:X DFT ising}
\end{figure}

We consider the non-integrable Ising chain with transverse and longitudinal fields and nearest-neighbor interactions. Its Hamiltonian is given by
\begin{equation}
    \label{eq:nonintetgrable_open}
    \begin{split}
    H(\lambda_t) = \lambda_t\sum_{j=1}^L\sigma_j^z + \sum_{j=1}^Lh_{j,x}\sigma_j^x &+ \sum_{j=1}^{L-1} J_j \sigma_j^z\sigma_{j+1}^z\\
    &+ h_1\sigma_1^z + h_L\sigma_L^z,
    \end{split}
\end{equation}
where $\sigma_j^{x,z}$ are the $x,z$ Pauli matrices for the $j$-th spin. The last two terms were added to break inversion symmetry \cite{ODonovan2025_QME_ETH}. We took $L=4$ for the numerical calculations that illustrate the fluctuation relations. The values of $J_j$ were drawn from a normal distribution with a mean of $J_m$ and a standard deviation of $0.1J_m$. The values of $h_{j,x}$ were drawn randomly from a normal distribution with a mean of $1.1J_m$ and a standard deviation of $0.1J_m$. We chose to randomly sample these values to ensure there are no eigenvalue crossings. Besides, we took $h_1=0.25J_m$ and $h_L=-0.25J_m$. The protocol $\lambda_t$ is taken as 
\begin{equation}
\label{eq:lambda_protocol ising}
\lambda_t=0.1J_m + 2.0J_mt/\tau,
\end{equation}
while the initial temperature was chosen as $\beta = J_m^{-1}$. Note that, due to the lack of symmetries in $H(\lambda)$, the energy spectrum is nondegenerate (this was checked numerically).

\begin{figure}[b!]
    \centering
    \includegraphics[width=0.91\linewidth]{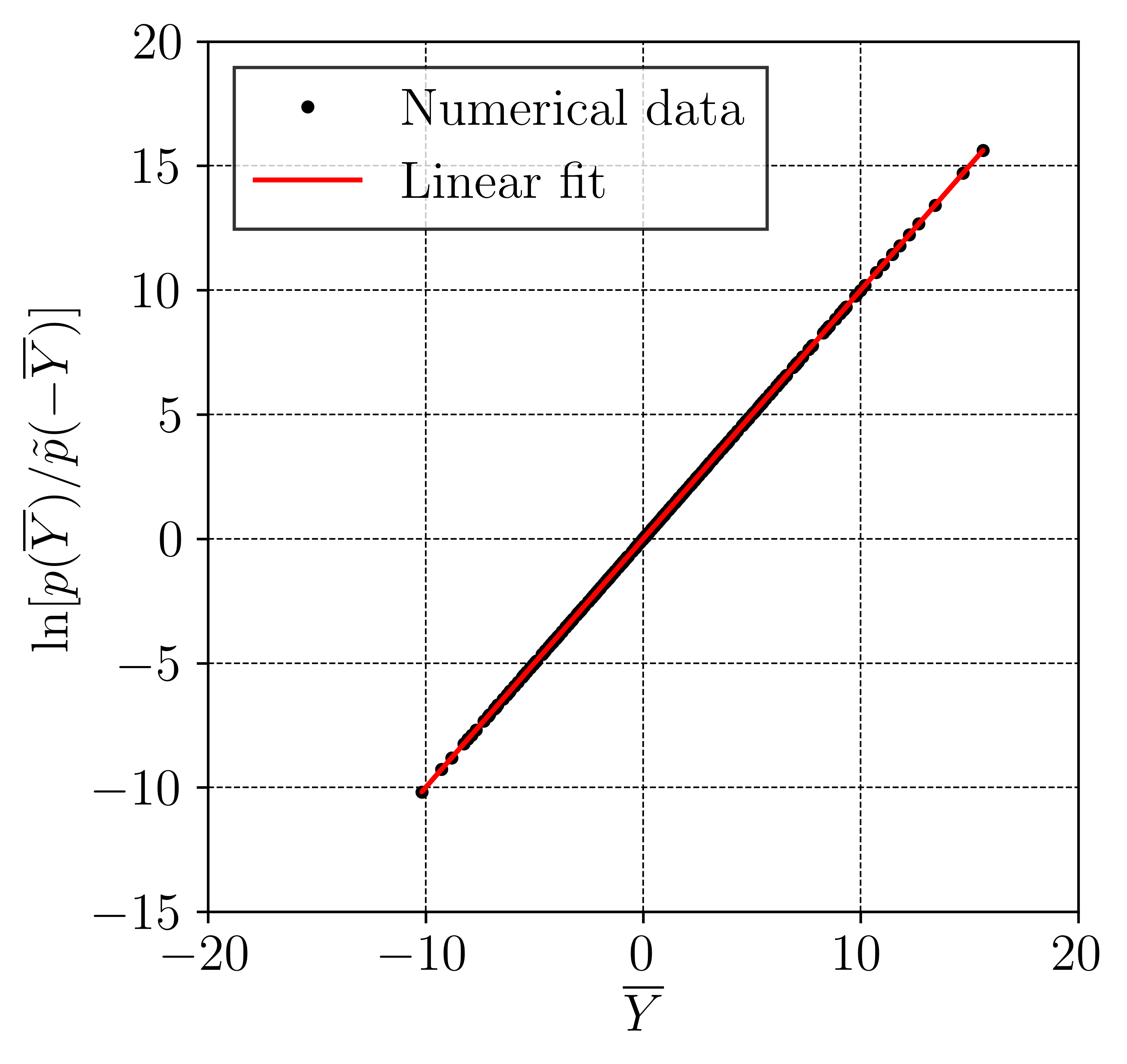}
    \caption{Numerical verification of the DFT for $Y$, Eq. (\ref{eq:Y_DFT}). The calculations were performed on the 4-spin Ising chain with the Hamiltonian of Eq. (\ref{eq:nonintetgrable_open}), protocol of Eq. (\ref{eq:lambda_protocol ising}), initial inverse temperature $\beta=J_m^{-1}$ and process duration $\tau=\hbar/J_m$.}
    \label{fig:Y DFT ising}
\end{figure}

In Figs.~\ref{fig:X DFT ising} and \ref{fig:Y DFT ising}, we show numerical verification of the two DFTs derived in this work for a process duration of $\tau =  \hbar/J_m$. The probabilities for $X$ were calculated directly from Eqs. (\ref{eq:p(m,n) def}) and (\ref{eq:p_tilde(n,m) def}). For Eq. (\ref{eq:p(m,n) def}), the term $U[P_n(\lambda_0)\rho(0)P_n(\lambda_0)]U^\dagger = \mu_n(\lambda_0)UP_n(\lambda_0)U^\dagger$ was calculated by numerically solving the Liouville equation
\begin{equation}
    \label{eq:liouville}
    \frac{d\rho}{dt}=-\frac{i}{\hbar}[H(\lambda_t), \rho(t)]
\end{equation}
with initial condition $\rho(0)=P_n(\lambda_0)$, thus obtaining $UP_n(\lambda_0)U^\dagger$. Similarly, we calculated the term $\tilde U [\tilde P_m(\tilde\lambda_0)\tilde\rho(0)\tilde P_m(\tilde\lambda_0)]\tilde U^\dagger = \mu_n(\lambda_0)\tilde U \tilde P_m(\tilde\lambda_0)\tilde U^\dagger$ by numerically solving the Liouville equation
\begin{equation}
    \label{eq:liouville_rev}
    \frac{d\tilde\rho}{d t}=-\frac{i}{\hbar}[\Theta H(\tilde \lambda_{ t})\Theta^\dagger, \tilde\rho( t)]
\end{equation}
with initial condition $\tilde \rho(0) = \tilde P_m(\tilde \lambda_0)$ and $\tilde\lambda_{t} = \lambda_{\tau - t}$. The probabilities of $\overline Y$ and $\tilde{\overline Y}$ were calculated analogously. The fourth-order adaptive Runge-Kutta method \cite{newman2013} was used to solve the systems of first-order differential equations implied by the Liouville equations (\ref{eq:liouville}) and (\ref{eq:liouville_rev}).

Both for Figs. \ref{fig:X DFT ising} and \ref{fig:Y DFT ising}, the angular coefficient $a$ of the linear fit agreed with the expected value, i.e., $a=\beta J_m$ for Fig. \ref{fig:X DFT ising} and $a=1$ for Fig. \ref{fig:Y DFT ising}, aside from a very small error (of the order of $10^{-6}a$) due to the numerical inaccuracy inherent in the computational solution of differential equations. 

\begin{figure}[b]
    \centering
    \includegraphics[width=0.85\linewidth]{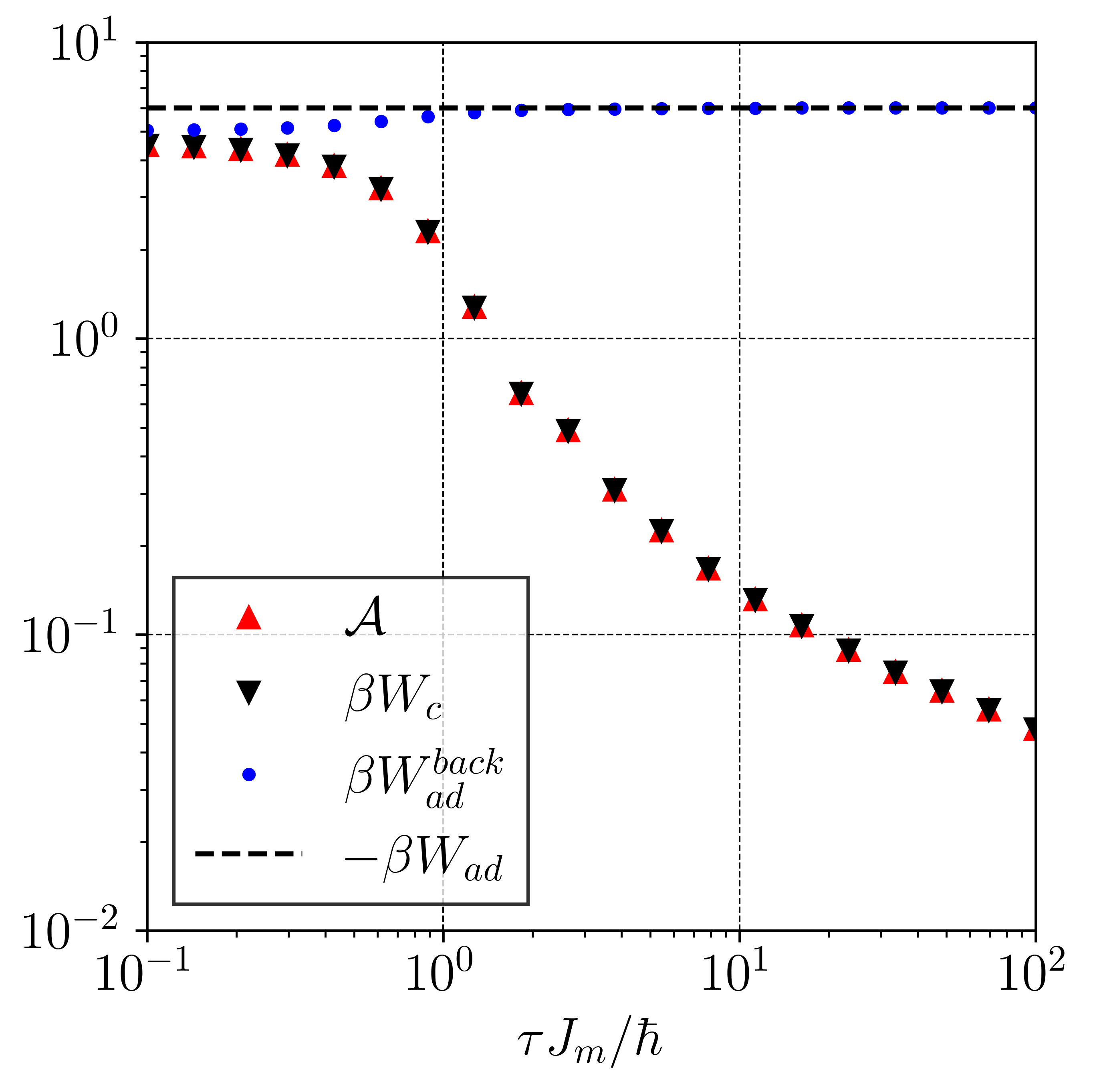}
    \caption{The quantities in Eq. (\ref{eq:W_c = D_ad}) as functions of the forward process duration $\tau$ for the 12-spin non-integrable Ising chain whose Hamiltonian is given in Eq. (\ref{eq:nonintetgrable_open}). The protocol for the variation of $\lambda$ is the one in Eq. (\ref{eq:lambda_protocol ising}) and the initial temperature is $\beta = J_m^{-1}$. We also provide a comparison between $W_\text{ad}^\text{back}$ and $-W_\text{ad}$.}
    \label{fig:W_c = D_ad ising}
\end{figure}

In Fig.~\ref{fig:W_c = D_ad ising}, we plot the quantities $\beta W_c$ and $\mathcal{A} = D[\rho(\tau)||\rho_\text{ad}(\lambda_\tau)]$ from Eq. (\ref{eq:W_c = D_ad}) as a function of the process duration $\tau$ for the linear protocol in Eq. (\ref{eq:lambda_protocol ising}), now for a larger chain with $L=12$. We also compare $W_\text{ad}^\text{back}$ and $-W_\text{ad}$. Let us briefly describe how all these quantities were calculated. For a given value of $\tau$, the density operator $\rho(\tau)$ at the end of the process was calculated by solving Eq. (\ref{eq:liouville}), with the initial condition $\rho(0)=\Pi_\beta(\lambda_0)$, using the fourth-order adaptive Runge-Kutta method. The adiabatic state $\rho_\text{ad}(\lambda_\tau)$ was calculated using the second line of Eq. (\ref{eq:rho_ad def}). To calculate $W_\text{ad}^\text{back}$, we first calculated $U_\text{ad}^\text{back}$, i.e., the evolution operator corresponding to the adiabatic process taking the external parameter from $\lambda_\tau$ to $\lambda_0$. This was done as follows. First, as the spectrum of the system at hand is nondegenerate, define the instantaneous energy eigenstates $\vert n(\lambda_t)\rangle$ by the eigenvalue equation $H(\lambda_t)\vert n(\lambda_t)\rangle = E_n(\lambda_t)\vert n(\lambda_t)\rangle$. Then, the adiabatic theorem yields
\begin{equation}
    U_\text{ad}^\text{back}\vert n(\lambda_\tau)\rangle = e^{i\phi_n^\text{back}}\vert n(\lambda_0)\rangle,
\end{equation}
where $\phi_n^\text{back}$ is the phase corresponding to the $n$-th eigenstate at the end of the backward adiabatic process and hence it is given by the geometric phase minus the dynamic phase \cite{sakurai_qm, messiah_qm_dover}. Therefore, we can write 
\begin{equation}
    U_\text{ad}^\text{back} = \sum_n e^{i\phi_n^\text{back}} \vert n(\lambda_0)\rangle\langle n(\lambda_\tau)\vert.
\end{equation}
Then $W_\text{ad}^\text{back}$ can be calculated using
\begin{equation}
    W_\text{ad}^\text{back} = \Tr\{ U_\text{ad}^\text{back} \rho(\tau)U_\text{ad}^{\text{back}\dagger} H(\lambda_0)\} - \Tr\{\rho(\tau)H(\lambda_\tau)\}.
\end{equation}
From this equation, one can see that $W_\text{ad}^\text{back}$ is independent of the phases $\phi_n^\text{back}$. Therefore, we did not calculate them, in order to save computing time.

We have just described how to obtain $\rho(\tau)$, $\rho_\text{ad}(\lambda_\tau)$, and $W_\text{ad}^\text{back}$, so that we can then calculate all the quantities in Fig.~\ref{fig:W_c = D_ad ising} for various process durations. Now, we discuss the results in Fig.~ \ref{fig:W_c = D_ad ising}. First, we should note that the values of $\mathcal{A}$ and $\beta W_c$ are in remarkable agreement, thus numerically confirming that $\beta W_c = \mathcal{A}$. Additionally, the comparison between $W_\text{ad}^\text{back}$ and $-W_\text{ad}$ shows that, although these quantities present similar values, it is notable that they are not equal. This is due to the fact that the initial state for calculating $W_\text{ad}^\text{back}$ is $\rho(\tau)$ and not $\rho_\text{ad}(\lambda_\tau)$. However, for $\tau$ sufficiently large, we see that $W_\text{ad}^\text{back} \rightarrow -W_\text{ad}$, as expected.

\begin{figure}[b!]
    \centering
    \includegraphics[width=0.9\linewidth]{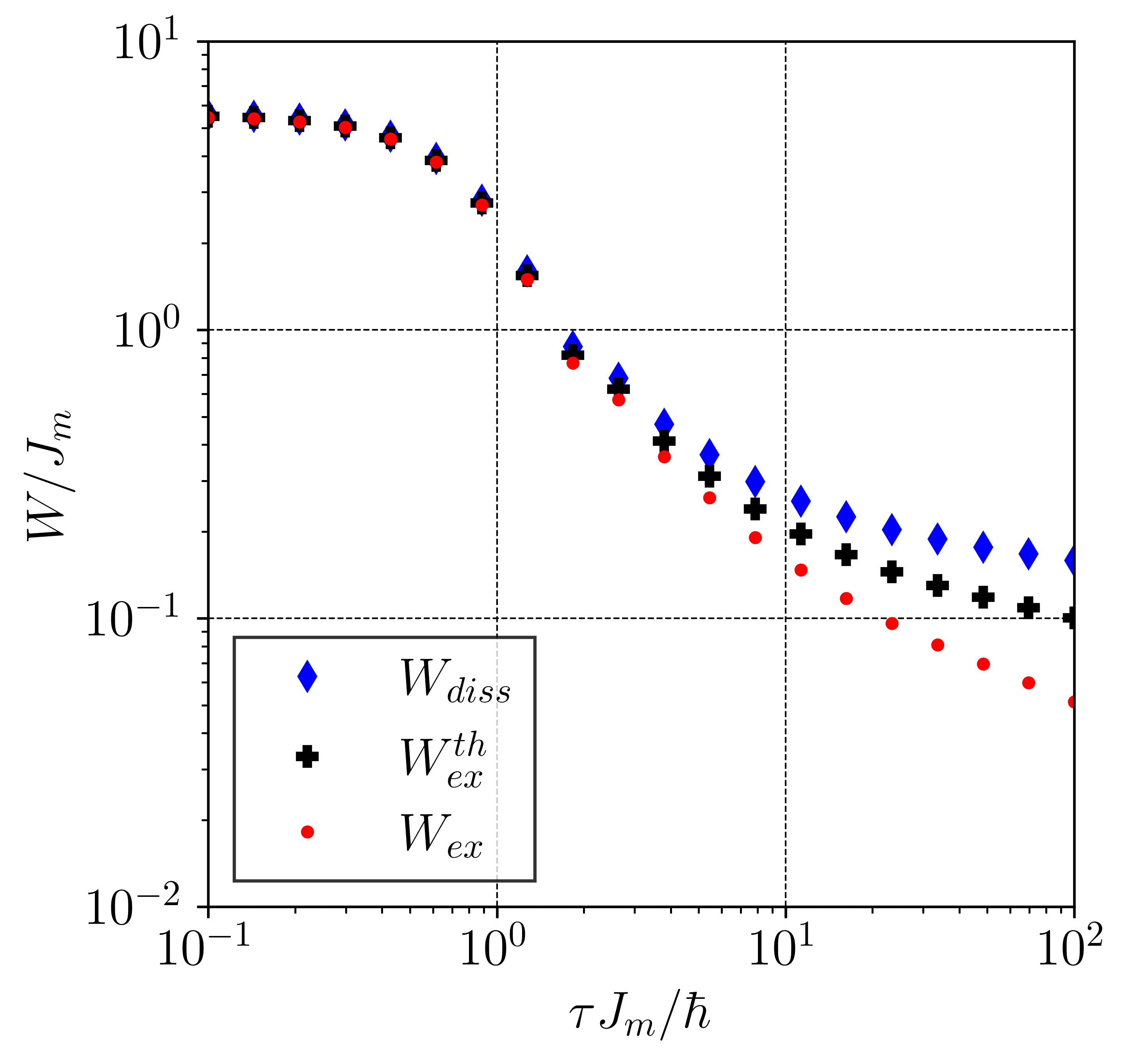}
    \caption{Illustration of Eq. (\ref{eq:inequality excess work}), where $W_\text{ex}=W-W_\text{ad}$, $W_\text{ex}^\text{th}=W-W_\text{ise}$, and $W_\text{diss}=W-\Delta F$ are plotted as functions of the process duration $\tau$ for the 12-spin non-integrable Ising chain whose Hamiltonian is given in Eq. (\ref{eq:nonintetgrable_open}). The protocol for the variation of $\lambda$ is the one in Eq. (\ref{eq:lambda_protocol ising}) and the initial temperature is $\beta = J_m^{-1}$.}
    \label{fig:inequality excess work ising}
\end{figure}

Figure \ref{fig:inequality excess work ising} compares the excess work $W_\text{ex}=W-W_\text{ad}$ with the thermodynamic excess work $W_\text{ex}^\text{th}=W-W_\text{ise}$. We also included $W_\text{diss}=W-\Delta F$ in the comparison, as this quantity was mentioned several times throughout this work. The chain size is $L=12$ once again. The graph confirms the inequalities in Eq. (\ref{eq:inequality excess work}).

Our results are clearly independent of the behavior of the system regarding (non-)integrability. However, to avoid possible vanishing transition probabilities often present in integrable systems due to selection rules, we chose a non-integrable Ising chain to illustrate our results.

\subsection{Systems with expanding/contracting spectra}

Now, we discuss the case of systems whose entire spectrum parametrically expands or contracts by the same factor during the protocol, that is
\begin{equation}
    \label{eq:spectral condition}
    E_n(\lambda_t) = f(\lambda_t)g(n), \forall n.
\end{equation}
Examples of systems that satisfy this condition are: any two-level system, the harmonic oscillator with a changing spring constant, the particle in a box with a changing length, and any system whose Hamiltonian is of the form $H(\lambda_t) = \lambda_tH_0$ (where $H_0$ is a time-independent observable). However, the condition in Eq. (\ref{eq:spectral condition}) is highly restrictive. Systems such as the transverse-field Ising chain with a time-dependent field do not satisfy it. 

Let us explore the consequences of Eq. (\ref{eq:spectral condition}). Combining it with Eq. (\ref{eq:X_and_Y_def}) yields
\begin{equation}
\begin{split}
    \label{eq:Y proportional to X}
    X &= E_{n_f}(\lambda_0) - E_{n_i}(\lambda_0)\\
    &= \frac{f(\lambda_0)}{f(\lambda_\tau)}[E_{n_f}(\lambda_\tau) - E_{n_i}(\lambda_\tau)] = \gamma Y,
\end{split}
\end{equation}
where we defined $\gamma \equiv f(\lambda_0)/f(\lambda_\tau)$. Thus, $Y$ is proportional to $X$ in the case considered here. As a consequence of Eq. (\ref{eq:X_and_Y_avgs}), we find 
\begin{equation}
    \label{eq:prop_1}
    \gamma \beta W_\text{ex} = \mathcal{A}.
\end{equation}
The factor $\gamma\beta$ is just $\overline \beta$, defined in Eq. (\ref{eq:beta_bar def}). Indeed, rewriting $\Pi_{\gamma\beta}(\lambda_\tau)$ by noting that $\gamma\beta E_n(\lambda_\tau) = \beta E_n(\lambda_0), \forall n$ yields
\begin{equation}
\label{eq:Pi_beta_bar = rho_ad}
\begin{split}
    \Pi_{\gamma\beta}(\lambda_\tau) &= \sum_n\frac{e^{-\gamma\beta E_n(\lambda_\tau)}}{Z_{\gamma\beta}(\lambda_\tau)}P_n(\lambda_\tau)\\
    &= \sum_n \frac{e^{-\beta E_n(\lambda_0)}}{Z_\beta (\lambda_0)} P_n(\lambda_\tau) = \rho_\text{ad}(\lambda_\tau).
\end{split}
\end{equation}
The adiabatic state is obtained through a unitary transformation of $\rho(0)$, so that $\Tr\{ \Pi_{\gamma\beta}(\lambda_\tau)\ln \Pi_{\gamma\beta}(\lambda_\tau)\} = \Tr\{ \rho(0)\ln\rho(0) \}$. By comparing this equation with the definition of $\overline \beta$, Eq. (\ref{eq:beta_bar def}), one concludes that $\gamma\beta = \overline \beta$. Thus, Eq. (\ref{eq:prop_1}) becomes
\begin{equation}
\label{eq:beta_bar Wex = D_ad}
    \overline\beta W_\text{ex} = \mathcal{A}\equiv D[\rho(\tau)||\rho_\text{ad}(\lambda_\tau)].
\end{equation}
Besides, since by Eq. (\ref{eq:Pi_beta_bar = rho_ad}) the adiabatic state is equal to a Gibbs state at temperature $\overline \beta$, one identifies $\overline \beta$ as the temperature at the end of the adiabatic evolution in this case.

The result in Eq. (\ref{eq:beta_bar Wex = D_ad}) is the same as Eq. (6) of \cite{Plastina2014_inner_friction}. We emphasize, however, that both we and the authors of \cite{Plastina2014_inner_friction} used the highly restrictive condition in Eq. (\ref{eq:spectral condition}), without which Eq. (\ref{eq:beta_bar Wex = D_ad}) is not valid in general. The more general result connecting work and non-adiabaticity is Eq. (\ref{eq:W_c = D_ad}).

As a final note, we point out that since $\rho_\text{ad}(\lambda_\tau) = \Pi_{\overline\beta}(\lambda_\tau)$ when Eq. (\ref{eq:spectral condition}) is satisfied, the difference between $W_\text{ad}$ and $W_\text{ise}$ vanishes (see Eq. (\ref{eq:W_ad-W_ise})), so that the thermodynamic minimal work principle becomes identical to the minimal work principle stated in \cite{AllahverdyanNieuwenhuizen2005_MinimalWorkPrinciple}. Thus, in this case, Eq. (\ref{eq:inequality excess work}) becomes
\begin{equation}
    \begin{split}
    &W_\text{ad} = W_\text{ise}\ge\Delta F\\
    \text{or} \quad &W_\text{ex} = W_\text{ex}^\text{th}\le W_\text{diss}.
    \end{split}
\end{equation}

\section{Conclusions and outlook}
\label{sec:conclusions}

In this work, we have derived detailed and integral fluctuation theorems for two quantities originally defined in Ref. \cite{Jarzynski2020} but whose physical interpretation was still lacking. A proper interpretation was provided not only to the corresponding fluctuation theorems but also to the average of the stochastic quantities themselves. They were shown to be related to the nonadiabaticity parameter $\mathcal{A}\equiv D[\rho(\tau)||\rho_\text{ad}(\lambda_\tau)]$ and the excess work $W_\text{ex}\equiv W-W_\text{ad}$, respectively. We then further investigated these quantities and derived additional results about the non-equilibrium thermodynamics of thermally isolated driven quantum systems. For instance, under the assumptions that the initial state is a Gibbs state and that the evolution is governed by the quantum Liouville equation, we have shown Eq. (\ref{eq:W_c = D_ad}), which relates the nonadiabaticity parameter and the work absorbed by the system during what we call the cyclic counterpart of the original process. We then argued that $\mathcal{A}$ is an interesting quantifier of irreversibility for thermally isolated driven systems. Concerning the excess work, we discussed and clarified its relation to the minimal work principle \cite{AllahverdyanNieuwenhuizen2005_MinimalWorkPrinciple}. Additionally, we used the new integral fluctuation theorem to derive what we call the thermodynamic minimal work principle, Eq. (\ref{eq:thermodynamic_minimal_work_principle}), which was shown to be the strongest bound attainable from the fluctuation theorems derived. We illustrated our results with calculations on a non-integrable Ising chain and on a relevant class of systems whose energy spectrum globally contracts or dilates. Such a class includes two-level systems, the quantum harmonic oscillator, and the particle in a three-dimensional box. 

We believe that our results open avenues for further research that we would like to comment next. First, a relevant question is whether there exists a suitable definition of non-equilibrium entropy such that its variation is equal (for thermally isolated systems starting from equilibrium) to the nonadiabaticity parameter. This definition of entropy should be such that, at equilibrium, it satisfies the same properties as the standard thermodynamic entropy under reasonable conditions, e.g., non-integrable systems in the thermodynamic limit. The existence of such quantity would make it so that the law of entropy increase would automatically be equivalent to Thomson's formulation of the Second Law.

Another question we believe deserves further investigation is related to the fact that, in standard Thermodynamics, reversible processes are said to take the system through a sequence of equilibrium states. In particular, the final state of a quasistatic transformation applied to a thermally isolated system initially in equilibrium would be an equilibrium state as well. However, as we have mentioned, adiabatic processes are reversible and, although $\rho_\text{ad}(\lambda_t)$ is a stationary state of $H(\lambda_t)$, it is in general not equal to one of the equilibrium ensembles from Statistical Mechanics, which describe equilibrium states. Thus, there are two apparently different predictions for the final state of a reversible process: the quantum mechanical prediction ($\rho_\text{ad}$) and the thermodynamic-statistical-mechanical prediction (corresponding to one of the equilibrium ensembles). This leads, for instance, to the difference between the two minimal work principles discussed in the text: $W_\text{ex}\ge0$ and $W_\text{ex}^\text{th}\ge0$. We intend to expand on this question in future work.

Additionally, a natural extension of the present work is to investigate how our results generalize in the presence of energy level crossings, which were assumed to be absent throughout this text. Because our analysis relies heavily on the adiabatic state $\rho_\text{ad}(\lambda_t)$, whose definition is generally tied to the absence of level crossings, such an extension may not be straightforward.

\begin{acknowledgments}
    J.V.M.S. and M.V.S.B. acknowledge financial support from FAPESP (Funda\c{c}\~ao de Amparo \`a Pesquisa do Estado de São Paulo), Grants No. 2023/16738-0 and 2025/07255-1. J.V.M.S. also acknowledges that this study was financed in part by the Coordenação de Aperfeiçoamento de Pessoal de Nível Superior - Brasil (CAPES) - Finance Code 001. We kindly thank P. Strasberg for insightful discussions.
\end{acknowledgments}

\appendix

\section{Principle of microreversibility}
\label{app:U_tilde}

Here, we prove Eq. (\ref{eq:micro_rev}). The proof is identical to the one in \cite{CampisiHanggiTalkner2011_Review}, with the exception that we do not require $[H(\lambda_t), \Theta]=0$, so the Hamiltonian that governs the evolution in the time-reversed process is $\Theta H(\tilde \lambda)\Theta^\dagger$. We begin by expressing $\tilde U(\tau-t, 0)$ as the application of subsequent infinitesimal evolution steps:
\begin{equation}
\label{eq:U_tilde derivation step 1}
\begin{split}
    \tilde U(\tau-t,0) = &\lim_{N\rightarrow\infty} e^{-i\varepsilon \Theta H(\tilde\lambda_{\tau-N\varepsilon})\Theta^\dagger/\hbar}\cdots e^{-i\varepsilon \Theta H(\tilde \lambda_\varepsilon)\Theta^\dagger/\hbar}\\
    &\times e^{-i\varepsilon \Theta H(\tilde \lambda_0)\Theta^\dagger/\hbar},
\end{split}
\end{equation}
where $\varepsilon=t/N$. Each exponential in the expression above can be rewritten as $e^{\Theta i\varepsilon H(\tilde \lambda)\Theta^\dagger/\hbar}$, as $\Theta$ is an anti-unitary operator and $\varepsilon\in \mathbb{R}$. Additionally, by writing the exponential as a Taylor series and using the fact that $\Theta^\dagger\Theta=\mathbb{1}$, one easily sees that $e^{\Theta i\varepsilon H(\tilde \lambda)\Theta^\dagger/\hbar} = \Theta e^{i\varepsilon H(\tilde\lambda)/\hbar}\Theta^\dagger$. Then, since $\tilde\lambda_{\tau-t} = \lambda_t$, Eq. (\ref{eq:U_tilde derivation step 1}) can be written as
\begin{equation}
\begin{split}
    \tilde U(\tau-t,0) = &\lim_{N\rightarrow\infty} \Theta e^{i\varepsilon H(\lambda_{N\varepsilon})/\hbar}\cdots e^{i\varepsilon  H(\lambda_{\tau-\varepsilon})/\hbar}\\
    &\times e^{i\varepsilon H( \lambda_\tau)/\hbar}\Theta^\dagger.
\end{split}
\end{equation}
Finally, as 
\begin{equation}
\begin{split}
    \Theta \lim_{N\rightarrow\infty}&e^{i\varepsilon H(\lambda_{N\varepsilon})/\hbar}\cdots e^{i\varepsilon H( \lambda_\tau)/\hbar}\Theta^\dagger\\
    &= \Theta\left[\lim_{N\rightarrow\infty} e^{-i\varepsilon H( \lambda_\tau)/\hbar} \cdots e^{-i\varepsilon H(\lambda_{N\varepsilon})/\hbar}  \right]^\dagger\Theta^\dagger\\
    &= \Theta U^\dagger(\tau, t) \Theta^\dagger,
\end{split}
\end{equation}
we obtain 
\begin{equation}
    \tilde U(\tau-t,0) = \Theta U^\dagger(\tau, t)\Theta^\dagger.
\end{equation}
which is Eq.~(\ref{eq:micro_rev}) and the proof is complete.

\section{Proof that $X=w_c$}
\label{app:X=w_c}

In order to prove that two random variables are equal, we must show that they take on the same values with the same probability distribution. Obviously, both $X$ and $w_c$ take on the same values $E_m(\lambda_0)-E_n(\lambda_0)$. Thus, it remains to show that $p(X=x)=p(w_c=x), \forall x$.

The probability distribution for $X$ is given in Eq. (\ref{eq:P}). Meanwhile, the probability distribution for $w_c$ is
\begin{equation}
    \label{eq:p(w_c)}
    p(w_c = x) = \sum_{mn} \delta[ x - (E_m(\lambda_0) - E_n(\lambda_0)) ]p_c(m, n),
\end{equation}
where $p_c(m,n)$ is the probability that the process described in steps 1 to 3 in Sec. \ref{sec:previous_results} starts at the eigenspace of $E_n(\lambda_0)$ and ends at the eigenspace of $E_m(\lambda_0)$. Then, to show that $X = w_c$, it suffices to show that $p(m,n) = p_c(m,n)$. Let us write $p_c(m,n)$ explicitly:
\begin{equation}
\label{eq:p_c_def}
    p_c(m,n) = \Tr\{ P_m(\lambda_0) U_\text{ad}^\text{back}U [\mu_n(\lambda_0) P_n(\lambda_0)] U^\dagger U_\text{ad}^{\text{back}\dagger} \},
\end{equation}
where we used the convention $U\equiv U(\tau,0)$ and defined $U_\text{ad}^\text{back}$ as the adiabatic evolution operator for step 3 of Sec.~\ref{sec:previous_results}. Since $U_\text{ad}^\text{back}P_m(\lambda_\tau)U_\text{ad}^{\text{back}\dagger} = P_m(\lambda_0)$ by the adiabatic theorem (see Eq. (\ref{eq:ad_theorem})), we have $U_\text{ad}^{\text{back}\dagger}P_m(\lambda_0)U_\text{ad}^\text{back} = P_m(\lambda_\tau)$. Thus, Eq. (\ref{eq:p_c_def}) becomes
\begin{equation}
    p_c(m,n) = \mu_n(\lambda_0) \Tr\{ P_m(\lambda_\tau)UP_n(\lambda_0)U^\dagger \} = p(m,n),
\end{equation}
where we used Eq. (\ref{eq:p(m,n)}) in the last equality. This completes the proof. Note that the proof was only possible because of the one-to-one mapping between the eigenspace with quantum number $n$ at the beginning and at the end of the backward adiabatic process. Thus, due to the fact that the backward process is adiabatic, performing an energy measurement at the end of the original process is equivalent to performing it at the end of the cycle.

\section{Entropy production and $W_\text{diss}$}
\label{app:W_diss and entropy production}

In this appendix, we provide the standard proof of the connection between the dissipated work $W_\text{diss}\equiv W-\Delta F$ and the entropy production $\Sigma$. We then point out the assumptions made throughout the proof and conclude that $W_\text{diss}$ cannot, in general, be rigorously identified with the entropy production if these conditions are not satisfied.

The proof of the proportionality between $W_\text{diss}$ and $\Sigma$ in a general setting\footnote{The configuration considered is a system in contact with a heat bath at temperature $T$.} is as follows.
\begin{equation}
    \begin{split}
    \Sigma &\equiv \Delta S_\text{universe} = \Delta S + \Delta S_\text{bath}\\
    &= \Delta S - Q/T\\
    &= T^{-1}(\Delta E-\Delta F) - T^{-1}(\Delta E-W)\\
    &= T^{-1}(W-\Delta F)
    \end{split}
\end{equation}
In the third line, we used the definition of the equilibrium Helmholtz free energy, $F = E - TS$, and the First Law of Thermodynamics, $\Delta E = W+Q$. 

This apparently rather innocent proof required several implicit assumptions, such as:
\begin{itemize}
    \item The entropy should be additive, so that we may write $\Delta S_\text{universe} = \Delta S + \Delta S_\text{bath}$. This need not be true in general, especially if the system is strongly coupled to the bath.
    \item The bath is ideal, so that we may write $\Delta S_\text{bath} = -Q/T$ and use the same $T$ throughout the proof.
    \item The system's initial and final states must be equilibrium states at temperature $T$, so that we may use $F_\text{eq} = E - TS$.
\end{itemize}
Since thermally isolated systems that are not put in contact with a heat bath by the end of the process generally do not achieve an equilibrium state with the same temperature as the initial one, we immediately see that the identification of $W_\text{diss}$ with the entropy production in this case is not justified.

\section{Global maximum of $-\langle \sigma'\rangle/\beta'$}
\label{app:global maximum}

In this appendix, we show that the point of maximum corresponding to $\beta' = \overline \beta$ is a global maximum of $-\langle \sigma'\rangle/\beta'$. Since $\beta' = \overline \beta$ is the only finite,  non-zero stationary point (see Eq. (\ref{eq:first_derivative})), it suffices to show that $\lim_{\beta'\rightarrow0}(-\langle \sigma'\rangle/\beta')$ and $\lim_{\beta'\rightarrow\infty}(-\langle \sigma'\rangle/\beta')$ are smaller than $-\langle \overline\sigma\rangle/\overline\beta$.

Let $d$ be the dimension of the Hilbert space. Then, $\beta'\rightarrow 0$ yields, by using Eq. (\ref{eq:-<sigma'>/beta'}),
\begin{equation}
\begin{split}
    \lim_{\beta'\rightarrow 0} -\frac{\langle \sigma'\rangle}{\beta'} &= \lim_{\beta'\rightarrow 0} \left[-\langle H(\lambda_\tau)\rangle_\text{ad} - \frac{1}{\beta'}\ln d + \frac{1}{\beta'}S_\beta( \lambda_0)\right]\\
    &= -\langle H(\lambda_\tau)\rangle_\text{ad} - \lim_{\beta'\rightarrow 0}\frac{1}{\beta'}[\ln d - S_\beta(\lambda_0)].
\end{split}
\end{equation}
Since $\ln d \ge S_\beta( \lambda_0)$, this limit yields $-\infty$. 

Now we discuss the limit $\beta'\rightarrow \infty$. In this limit, the last term of Eq. (\ref{eq:-<sigma'>/beta'}) approaches zero. On the other hand, the term $-\beta'^{-1}\ln Z_{\beta'}(\lambda_\tau)$ approaches $E_0(\lambda_\tau)$, the ground state of the final Hamiltonian, as the only relevant term in $Z_{\beta'}(\lambda_\tau) = \sum_n \Tr\{ P_n(\lambda_\tau) \} e^{-\beta'E_n(\lambda_\tau)}$ is the one for $n=0$ in the limit $\beta'\rightarrow \infty$. Then,
\begin{equation}
\begin{split}
    \lim_{\beta'\rightarrow\infty} -&\frac{\langle \sigma'\rangle}{\beta'} = -\langle H(\lambda_\tau)\rangle_\text{ad} + E_0(\lambda_\tau)\\
    &\le -\Tr\{ \rho_\text{ad}(\lambda_\tau) H(\lambda_\tau) - \Pi_{\overline\beta}(\lambda_\tau)H(\lambda_\tau) \}\\
    &= -\overline\beta^{-1}D[\rho_\text{ad}(\lambda_\tau)||\Pi_{\overline\beta}(\lambda_\tau)],
\end{split}
\end{equation}

\noindent where we used Eq. (\ref{eq:W_ad-W_ise}) in the last line.
It follows that the limit $\beta'\rightarrow \infty$ also yields a smaller value of $-\langle \sigma'\rangle/\beta'$ than $\overline\beta$. 

We conclude that the sharpest bound on the work attainable from this IFT is the one in Eqs. (\ref{eq:W_th>0 from IFT}) and (\ref{eq:lower bound}).

%\vspace{3 cm}

\bibliography{apssamp}% Produces the bibliography via BibTeX.

\end{document}